\documentclass{article}
\usepackage{paper}

\title{Bayesian shrinkage priors for penalized synthetic control estimators in the presence of spillovers}
\author[1,*]{Esteban Fern\'andez-Morales}
\author[1]{Arman Oganisian}
\author[1]{Youjin Lee}
\affil[1]{Department of Biostatistics, Brown University, Rhode Island, United States}
\affil[*]{\textit{email:} esteban.fernandez@brown.edu}
\date{\ }

\makeatletter\let\Title\@title\makeatother
\newcommand{\revision}[1]{\textcolor{black}{#1}}

\begin{document}

\maketitle

\begin{abstract}
Synthetic control (SC) methods are widely used to estimate the effects of policy interventions, especially those targeting specific geographic regions, referred to as units. 
These methods construct a weighted combination of untreated units, forming a ``synthetic'' control that approximates the counterfactual outcomes of the treated unit had the intervention not occurred. 
\revision{Although neighboring areas are often selected as controls due to their similarity in observed and unobserved characteristics, their proximity can lead to spillover effects, where the intervention indirectly impacts control units, potentially biasing causal estimates.}
To address this challenge, we introduce a Bayesian SC framework with \revision{utility-based} shrinkage priors. 
Our approach extends traditional penalization techniques (i.e., horseshoe, spike-and-slab) by incorporating a \revision{utility function} that combines \revision{covariate similarity} and \revision{spatial distance}.
\revision{This provides a metric that guides the data-driven selection of control units based on their relevance and spillover risk, which is assumed to increase with spatial proximity.}
Rather than outright excluding neighboring units, the method balances bias and variance by reducing the importance of potentially contaminated controls by spillovers.
We evaluate the proposed method through simulation studies at varying spillover levels and apply it to assess the impact of Philadelphia's 2017 beverage tax on the sales of sugar-sweetened and artificially sweetened beverages in mass merchandise stores.
\end{abstract}

\section{Introduction}
\label{sec:intro}

Evaluating the impact of policy interventions remains a central goal in economics and public health.
However, policies often have unintended effects that extend beyond their intended targets. 
For example, Philadelphia's excise tax on sugar-sweetened and artificially sweetened beverages~(SSB) reduced purchases within the city, but coincided with increases in neighboring areas, a likely sign of cross-border shopping~\citep{roberto_association_2019}. 
Similar patterns have been documented in classical applications, such as Germany's reunification and California's tobacco control program, where nearby areas experienced unintended spillover effects despite not being directly targeted.

These spillover effects pose challenges for causal inference, particularly when neighboring units -- commonly used as controls due to demographic and socioeconomic similarities with the treated -- are themselves indirectly affected. 
If included in the analysis, such controls no longer represent valid counterfactuals (e.g., what the treated area's outcome would have been without the intervention), potentially leading to biased estimates of the treatment effect. 
This issue arises in both the difference-in-differences~(DiD)~\citep{ashenfelter_estimating_1978} and synthetic control~(SC)~\citep{abadie_economic_2003} methods, which rely on the assumption that control units remain unaffected by the treatment.
Throughout the manuscript, we use the term ``treatment'' interchangeably with ``intervention'' or ``policy,'' and ``spillover'' with ``contamination'' or ``interference.''

\subsection{\revision{Causal inference for policy evaluation}}

The DiD framework estimates a treatment effect by comparing changes in outcomes between treated and control groups over time~\citep{ashenfelter_estimating_1978, ashenfelter_using_1984}. 
Its validity rests on the parallel trends assumption: that treated and control units would have followed similar trajectories in the absence of treatment. 
The SC method constructs a weighted combination of control units, namely a donor pool, to approximate the treated unit's counterfactual outcomes~\citep{abadie_synthetic_2010}.
The treatment effect is then estimated by taking the difference between the observed and synthetic outcomes of the treated unit.

While both methods are widely used, they break down in the presence of spillovers. 
DiD becomes biased as outcome trends no longer reflect what would have occurred in the absence of the intervention. 
SC may assign positive weights to contaminated controls, distorting the counterfactual and biasing the estimated treatment effect.
As a result, the need for methods that can mitigate or correct for spillover effects has led to several innovations.

DiD methods have handled interference using exposure mappings, which quantify the degree of indirect exposure or spillover experienced by each unit~\citep{verbitsky-savitz_causal_2012, hettinger_estimation_2023, lee_policy_2023}.
Recent approaches in the SC framework have addressed spillovers using exclusion rules or bias correction adjustments. 
For example, \citet{grossi_direct_2024} and \citet{marinello_impact_2021} propose excluding geographically adjacent controls from the donor pool to limit contamination.
However, this may fail to capture units that, while not directly adjacent, are nonetheless indirectly affected by the intervention.
Others, such as \citet{cao_estimation_2019} and \citet{di_stefano_inclusive_2024}, address spillover bias by introducing systems of equations or transformation matrices to jointly estimate direct and indirect effects. 
Likewise, these methods rely on non-singularity conditions, which, if violated, render the system unsolvable.

\subsection{\revision{A motivating application}}

\revision{We illustrate our motivating example of Philadelphia's 2017 beverage tax.
This excise tax, implemented on January~1,~2017, was aimed at reducing the consumption of SSB by reducing its sales volume throughout the city.
In evaluating the effect of the tax on SSB sales in Philadelphia, a central challenge is cross-border shopping, where residents travel to neighboring untaxed areas to purchase beverages, increasing sales in these regions, and confounding the true impact of the tax~\citep{roberto_association_2019,hettinger_doubly_robust_estimation_2025}.
If such neighboring units are included as controls, the treatment effect may be overestimated, as the decline in sales for Philadelphia is offset by increases just outside the city.}

\begin{figure}[t]
    \centering
    \includegraphics[width=0.6\linewidth]{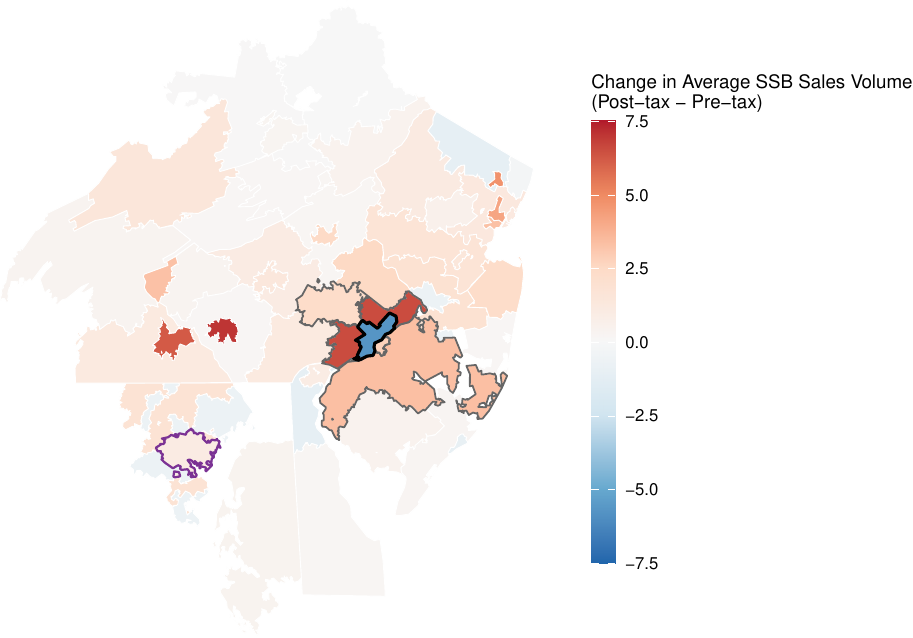}
    \caption{\revision{Change in post- versus pre-tax SSB sales volume by ZIP3 region. Differences in sales volume are measured in units of 10,000 fluid ounces from mass merchandise stores within each region. Negative values indicate a decline in sales after the excise tax. Colored borders highlight key areas: Philadelphia, the treated unit (black); its neighboring regions (gray); and Baltimore, the control unit with the closest covariate profile (purple), emphasized for its similarity but lower risk of spillover from cross-border shopping. Covariate similarity is calculated using the utility function $u_C$ in~\eqref{eq:utility-function} with $\kappa_d = 1$.}}
    \label{fig:change-sales-volume}
\end{figure}

\revision{Figure~\ref{fig:change-sales-volume} illustrates evidence of spillover effects, whose degree depends on the spatial proximity to Philadelphia.
It shows that post-tax SSB sales increase in three-digit zip code (ZIP3) regions that border Philadelphia, while decreasing within the city.
If these areas substantially contribute to the SC due to their similar characteristics to the treated area, the estimated effect may overstate the impact of the tax by offsetting the decline in sales with a shift in purchases to nearby untaxed regions.
In contrast, a more distant region like Baltimore (purple border) shows a covariate profile similar to that of Philadelphia without evident signs of spillover, making it a more suitable candidate for the donor pool.
This application motivates the need for a method that jointly considers both covariate similarity and geographic distance when selecting control units. 
Therefore, contributions are determined in a way that balances similarity with the risk of spillover.}

\subsection{\revision{Our contributions}}

To determine a control unit's contribution to the SC while accounting for two distinct and somewhat contradictory factors, we propose a Bayesian SC framework that incorporates shrinkage priors informed by a utility function that balances covariate similarity and spatial distance.
Our approach incorporates covariate and spatial information into donor pool selection, avoiding ad hoc exclusion rules. 
Specifically, the proposed utility function guides the amount of shrinkage applied to each control unit, reducing the importance of likely spillover-affected units while prioritizing those that closely resemble the treated unit.
\revision{We adopt the SC framework, rather than alternatives such as DiD, because the data consist of many potential controls but only one treated unit. 
This setting is well suited for constructing a weighted combination of controls to approximate counterfactual outcomes, without invoking or testing the parallel trends assumption.}
\revision{Furthermore, we extend the literature on Philadelphia's beverage tax policy~\citep{cawley_impact_2019,roberto_association_2019,lee_policy_2023,hettinger_doubly_robust_estimation_2025} by evaluating how incorporating covariate similarity and spatial distance in the selection of controls influences treatment effect estimates. 
Lastly, we demonstrate the effectiveness of our proposed approach by comparing it to several current and standard SC methods through extensive simulation studies under different levels of spillover.}

The remainder of this article proceeds as follows. 
Section~\ref{sec:framework} introduces the notation, causal estimand, and its underlying assumptions. 
Section~\ref{sec:methods} presents our Bayesian model and details the proposed utility-based shrinkage priors.
The simulation studies in Section~\ref{sec:simulations} assess their finite sample performance against numerous alternatives. 
Section~\ref{sec:application} applies the method to estimate the impact of the beverage tax in Philadelphia. 
We conclude in Section~\ref{sec:discussion} with limitations and directions for future research.

\section{Notations, Assumptions, and Causal Estimands}
\label{sec:framework}

We consider aggregated units such as cities, states, regions, or other large population areas.
\revision{For example, in our data application, units are defined by ZIP3 regions, where a single treated unit (Philadelphia) is observed, and the intervention (beverage tax) can be uniquely assigned.}
Let $i \in \qty[n]$ index units and $t \in \qty[T]$ index discrete time points, where $\qty[m] \coloneqq \qty{1, \ldots, m}$ denotes the set of integers from $1$ to $m$. 
The primary outcome is denoted by $Y_{it} \in \R$ for each unit $i$ at time $t$.
\revision{Only one unit ($i = 1$) is treated from time $T_0$ onward, where $T_0 < T$ denotes the number of pre-intervention periods.}
The remaining $J \coloneqq n - 1$ control units form the donor pool, with the collection of their outcomes at time $t$ denoted by $\bsV_t = (Y_{2t}, \ldots, Y_{nt})$.
Let $Z_{it} \in \qty{0,1}$ indicate whether unit $i$ receives the treatment directly at time $t$.
This treatment is binary, assigned once, and irreversible: $Z_{it} \leq Z_{is}$ for $t \leq s$.
The treatment vector for all $n$ units at time $t$ is $\bsZ_t = (Z_{1t}, \ldots, Z_{nt})$.

\revision{For each unit $i$, we observe baseline covariates $\bsX_i \in \R^q$ and spatial coordinates $\bsP_i \in \R^k$, where $q$ and $k$ denote their respective dimensions.}
For example, in our data application, $k = 2$ corresponds to the latitude and longitude of the geographic centroid for each ZIP3 region.
\revision{The combined feature vector is denoted by $\bsW_i = (\bsX_i, \bsP_i)$, with the collection of features for the sample defined as $\bsW = (\bsW_1,  \ldots, \bsW_n)$.}
\revision{We define the data unaffected by the intervention as $\bsD^o = (\bsY^o, \bsZ, \bsV, \bsW)$, where $\bsY^o = \qty{Y_{1t} \colon t \leq T_0}$ denotes pre-intervention outcomes of the treated unit, $\bsZ = (\bsZ_1, \ldots, \bsZ_T)$ the treatment assignments, and $\bsV = (\bsV_1, \ldots, \bsV_T)$ the control outcomes.}
\revision{Before treatment, $\bsZ_t = (0, \mathbf{0}_J)$; after treatment, $\bsZ_t = (1, \mathbf{0}_J)$, where $\mathbf{0}_J$ is a length-$J$ vector of zeros.}
For simplicity, we denote the history and future of a variable $A$ as $\obar{A}_t = (A_1, \ldots, A_t)$ and $\ubar{A}_t = (A_t, \ldots, A_T)$.

We adopt the potential outcomes framework~\citep{rubin_estimating_1974}, extending it to accommodate interference. 
\revision{Let $Y_{1t}(z_{1t}, \mathbf{0}_J)$ denote the potential outcome for the treated unit at time~$t$ under treatment assignment $z_{1t}$, assuming that no control unit is directly treated ($z_{it} = 0$ for all $i \neq 1$).}
Our target estimand is the causal effect of the intervention on the treated unit at time $t > T_0$:
\begin{equation}
    \label{eq:target-estimand}
    \tau_t \coloneqq \revision{Y_{1t}(1, \mathbf{0}_J)} - \revision{Y_{1t}(0, \mathbf{0}_J)}.
\end{equation}
Since \revision{$Y_{1t}(1, \mathbf{0}_J)$} is observed for $t > T_0$, inference focuses on imputing the counterfactual \revision{$Y_{1t}(0, \mathbf{0}_J)$} using pre-intervention data and post-intervention control outcomes.

\revision{Let $\bsS_t = (\bsV_t, Y_{1(t-1)}(0, \mathbf{0}_J))$ denote the state vector at time~$t$, containing current control outcomes and the lagged treated unit's outcome.
For $t = 1$, we set $\bsS_1 \equiv \bsV_1$, since no lagged outcome is available.}
To identify $\tau_t$, we require the following assumptions:
\begin{enumerate}
    \item[(A1)] \revision{Consistency: if $Z_{1t} = z_{1t}$, then $Y_{1t} = Y_{1t}(z_{it}, \mathbf{0}_J)$.}
    \item[(A2)] \revision{Markov property: $Y_{1t}(0, \mathbf{0}_J) \indep \obar{\bsS}_{t-1} \mid \bsS_t, \bsX_1$.}
    \item[(A3)] \revision{Sequential ignorability: $\ubar{Y}_{1t}(0, \mathbf{0}_J) \indep \bsZ_t \mid \bsS_t, \bsX_1$.}
\end{enumerate}

\revision{These assumptions link observed data with unobserved counterfactuals.
Assumption~1 ensures observed outcomes correspond to well-defined potential outcomes under the observed assignment.
Assumption~2 introduces a first-order Markov assumption that simplifies the dependence structure between the potential outcome and its observed history.
This structure is more flexible than traditional SC methods as it can accommodate higher-order temporal dependencies.
Assumption~3 generalizes the usual sequential ignorability assumption to interference settings, such that treatment assignments are assumed to be independent of the treated unit's counterfactual outcomes, given the observed history.
This assumes a sufficient covariate adjustment to capture relevant confounders.}

\revision{We adopt a missing data perspective~\citep{rubin_bayesian_inference_causal_1978}, treating the unobserved outcomes $\bsY^m = \qty{Y^m_{1t}(0, \mathbf{0}_J) \colon t > T_0}$ as missing and imputing them using their posterior predictive distribution (PPD), conditional on $\bsD^o$.}
Let $\bstheta \in \Theta$ denote the model parameters for the untreated outcome process, with $\Theta$ representing the parameter space and $\bstheta$ assumed to be time-invariant.
\revision{We further let $\bsgamma \in \Gamma$ denote the hyperparameters that govern the prior distribution of $\bstheta$, forming the basis for the data-driven hyperpriors, the details of which are described in Section~\ref{sub:shrinkage-priors}.}
\revision{Under Assumptions~1-3, we define the PPD~as
\begin{equation}
    \label{eq:posterior-predictive-distribution}
    \begin{aligned}
        & p(\bsY^m \mid \bsD^o) \\
        & \propto \int_\Theta \int_\Gamma \prod_{t > T_0} p(Y_{1t}^m(0, \mathbf{0}_J) \mid \bsS_t, \bsX_1, \bstheta) \\
        & \quad \times \prod_{t \leq T_0} p(Y_{1t}(0, \mathbf{0}_J) \mid \bsS_t, \bsX_1, \bstheta) p(\bstheta \mid \bsgamma) p(\bsgamma \mid \bsW) \, \dd \bsgamma \, \dd \bstheta.
    \end{aligned}
\end{equation}
where the conditional state vector $\bsS_t$ includes the previously imputed outcome $Y_{1(t-1)}^m(0, \mathbf{0}_J)$ for $t > T_0 + 1$.}

In practice, posterior draws of $\bstheta$ and imputed values for $\bsY^m$ are obtained using Monte Carlo methods; the details are provided in Section~\ref{sub:posterior-sampling}.
This enables the estimation of $\tau_t$ for each $t > T_0$.
\revision{In what follows, we introduce all identifiable components that define the PPD: the outcome model $p(Y_{1t}(0, \mathbf{0}_J) \mid \bsS_t, \bsX_1, \bstheta)$, the prior distribution $p(\bstheta \mid \bsgamma)$, and the data-dependent hyperprior $p(\bsgamma \mid \bsW)$.}
\revision{The derivation of the PPD in~\eqref{eq:posterior-predictive-distribution} and additional assumptions are provided in Web~Appendix~A.}

\section{Methods}
\label{sec:methods}

\revision{Bayesian SC models typically define the potential outcomes of the treated unit under no intervention as a function of observed covariates and control units' pre-intervention outcomes~\citep{brodersen_inferring_2015, kim_bayesian_2020, pang_bayesian_2022}.
Building on this, we specify the following for each $t \in \qty[T]$:
\begin{equation}
    \label{eq:mean-specification}
    \begin{aligned}
        & \E{Y_{1t}(0, \mathbf{0}_J)}{\bsS_t, \bsX_1, \bseta} \\
        & \quad = \bsX_1' \bsvartheta + \bsV_t' \bsbeta + \varphi Y_{1(t-1)}(0, \mathbf{0}_J) \\
        & \quad \eqqcolon m(\bsS_t, \bsX_1; \bseta),
    \end{aligned}
\end{equation}
where $\bseta = (\bsvartheta, \bsbeta, \varphi)$, with $\bsvartheta \in \R^q$ denoting the covariate effects, $\bsbeta \in \R^J$ the SC coefficients, and $\varphi \in \R$ the autoregressive coefficient.
We assume a normal model:
\begin{equation}
    \label{eq:outcome-model}
    Y_{1t}(0, \mathbf{0}_J) \mid \bsS_t, \bsX_1, \bseta, \phi \sim \Normal(m(\bsS_t, \bsX_1; \bseta), \phi),
\end{equation}
where $m(\cdot)$ dictates the mean and $\phi \in \R^+$ denotes the variance of the outcome.
Let $\bstheta = (\bseta, \phi)$ denote the model parameters.
The history $\bsS_t$ captures the dependence on current control outcomes $\bsV_t$ and its most recent lagged outcome $Y_{1(t-1)}(0, \mathbf{0}_J)$, which can be imputed or observed depending on time $t$.}

The coefficients $\bsbeta = (\beta_2, \ldots, \beta_n)$ function similarly to SC weights, forming a linear combination $\bsV_t' \bsbeta$ of the control outcomes.
\revision{However, unlike the standard SC, we allow $\bsbeta$ to lie outside the simplex (convex hull).
This relaxation can improve pre-intervention fit, particularly when the characteristics of the donor pool are very different from the treated unit~\citep{doudchenko_balancing_2016,ben-michael_augmented_synthetic_control_2021,athey_matrix_completion_methods_2021}.
The trade-off is reduced interpretability, since the coefficients in $\bsbeta$ no longer sum to one and instead span the full real line.}

\revision{Full Bayesian inference requires prior specifications for the model parameters in $\bstheta$.
In this work, we focus on a class of utility-based shrinkage priors for $\bsbeta$.
We modify two widely used shrinkage priors -- the horseshoe and spike-and-slab -- into what we call the distance horseshoe (DHS) and distance spike-and-slab (DS2) priors.
These priors are governed by the outcome variance $\phi$ and a set of data-dependent hyperparameters $\bsgamma$ in~\eqref{eq:posterior-predictive-distribution}. 
Both DHS and DS2 adjust the amount of shrinkage applied to each element in $\bsbeta$ -- that is, each control unit's contribution to the SC -- based on its covariate similarity to and spatial distance from the treated unit.
Notably, $\bstheta$ and $\bsgamma$ represent distinct components: while $\bstheta$ governs the outcome distribution, $\bsgamma$ defines the structure of the prior on $\bsbeta$.
A complete derivation of the posterior distribution, along with prior specifications for the remaining parameters ($\bsvartheta$, $\varphi$, and $\phi$) is provided in Web~Appendix~B.}

\subsection{\revision{Utility function}}

\revision{To control the shrinkage applied to each SC coefficient in $\bsbeta$, we define a utility function that combines covariate similarity and spatial distance to the treated unit.
Although both covariates and spatial coordinates are included in $\bsW_i = (\bsX_i, \bsP_i)$, they serve different roles: similarities in $\bsX_i$ and $\bsX_1$ help reduce confounding, while the distance between $\bsP_i$ and $\bsP_1$ may indicate the potential for spillover bias.
However, a common challenge is that neighboring units often resemble the treated unit in both covariates and spatial proximity, making it difficult to account for both sources of bias simultaneously.}

\revision{To balance these competing factors, we define the utility of unit $i \in [n] \setminus \qty{1}$ relative to the treated unit as
\begin{equation}
    \label{eq:utility-function}
    u_C(\bsW_i, \bsW_1)
        = \kappa_d d_X(\bsX_i, \bsX_1) + (1 - \kappa_d) d_P(\bsP_i, \bsP_1),
\end{equation}
where $\kappa_d \in \qty[0,1]$ is an importance weight that controls the trade-off between covariate similarity~$d_X$ and spatial distance~$d_P$.
The measures $d_X$ and $d_P$ are specified as
\begin{align*}
    d_X(\bsX_i, \bsX_1) 
        &= 1 / \qty(1 + \norm{\bsX_i - \bsX_1}), \\
    d_P(\bsP_i, \bsP_1) 
        &= \norm{\bsP_i - \bsP_1} \big/ K,
\end{align*}
where $\norm{\,\cdot\,}$ denotes the Euclidean norm and $K \in \R^+$ is a scaling factor reflecting the maximum observed distance between units.
Both $d_X$ and $d_P$ are normalized to the unit interval $[0, 1]$.}

We use the Euclidean distance because of its general applicability.
The covariates should be standardized to the unit mean and scale, with spatial locations defined by the units' centroids, making the Euclidean norm an appropriate choice for measuring both covariate similarity and spatial distance.
\revision{Alternative similarity measures include the Jaccard index~\citep{jaccard_distribution_1912} for binary and categorical covariates, the Mahalanobis distance~\citep{mahalanobis_generalized_1936} to account for correlations between covariates, or a weighted Euclidean norm that assigns greater importance to certain covariates.}

When $\kappa_d = 1$, the utility function in~\eqref{eq:utility-function} considers only covariate similarity in SC selection to reduce confounding; while $\kappa_d = 0$ considers only spatial distance to mitigate spillover bias. 
Both factors are considered for $\kappa_d \in (0, 1)$, with a greater emphasis on covariates as $\kappa_d$ approaches one.
\revision{Given $\kappa_d \in [0, 1]$, the utility score $u^C_{i,1} \coloneqq u_C(\bsW_i, \bsW_1)$ quantifies the relative ``usefulness'' of each control unit to construct the SC.
We outline its incorporation into the DHS and DS2 as follows.}

\subsection{\revision{Defining the utility-based priors}}
\label{sub:shrinkage-priors}

We propose the DHS by modifying the classical horseshoe prior~\citep{carvalho_horseshoe_2010}, which is well suited to sparse settings where $T_0 < J$. 
For each $i \in \qty[n] \setminus \qty{1}$, the prior on $\beta_i$ is given by
\begin{equation}
    \label{eq:distance-horseshoe}
    \begin{gathered}
        \beta_i \mid \phi, \lambda_i, \zeta \sim \Normal(0, \phi \lambda_i^2 \zeta^2), \\
        \lambda_i \mid u^C_{i,1} \sim \HC(0, u^C_{i,1}),
    \end{gathered}
\end{equation}
where $\zeta \in \R^+$ is the global shrinkage parameter shared between units, $\lambda_i \in \R^+$ is a local parameter that allows unit-specific variation, and $\phi$ is the outcome variance from~\eqref{eq:outcome-model}.
\revision{The hyperparameters are $\bsgamma = (\lambda_2, \ldots, \lambda_n, \zeta)$, where only $\lambda_i$ depends on $\bsW$ via $u^C_{i,1}$.}
\revision{Smaller utility scores induce stronger shrinkage ($\beta_i \to 0$) by reducing the scale of $\lambda_i$, thereby discouraging the inclusion of controls with low covariate similarity or a close spatial proximity to the treated unit, depending on $\kappa_d$.}

Alternatively, we propose DS2 by adapting the standard spike-and-slab mixture prior~\citep{mitchell_bayesian_1988, george_variable_1993} to incorporate a \revision{utility}-based inclusion mechanism.
\revision{For each $i \in \qty[n] \setminus \qty{1}$,
\begin{equation}
    \label{eq:distance-spike-and-slab}
    \begin{gathered}
        \beta_i \mid \omega_i, \phi, \nu \sim (1 - \omega_i) \delta_0 + \omega_i \Normal(0, \phi \nu^2), \\
        \omega_i \coloneqq \one(\revision{u^C_{i,1}} > \rho),
    \end{gathered}
\end{equation}
where $\omega_i \in \qty{0,1}$ indicates the assignment of the mixture component, $\nu \in \R^+$ controls the variance for the normal distribution (the slab), $\delta_0$ is a point mass at zero (the spike), $\one(\cdot)$ is the indicator function, and $\rho \in [0,1]$ is a user-specified cutoff.}
\revision{The hyperparameters are $\bsgamma = (\omega_2, \ldots, \omega_n, \nu)$, with $\omega_i$ obtained deterministically by comparing $u^C_{i,1}$ to $\rho$.}
When utility is based solely on spatial distance ($\kappa_d = 0$), the threshold $\rho$ determines the minimum distance a control unit must be from the treated unit for it to be included in the donor pool.
This formulation separates units into those allowed to contribute ($\beta_i \neq 0$) and those strictly excluded from the SC ($\beta_i = 0$), based on \revision{covariate similarity}, \revision{spatial distance}, or both.
\revision{We include a summary table of notation, graphical representations of the prior structures, and additional technical details -- such as prior specifications for $\zeta$ and $\nu$ -- in Web~Appendix~B.}

\subsection{Posterior sampling algorithm}
\label{sub:posterior-sampling}

Given the model and prior specifications, we draw samples from the posterior distribution of $\bstheta$ using Markov chain Monte Carlo methods.
\revision{To generate $R$ posterior draws, we use Hamiltonian Monte Carlo~\citep{homan_nouturn_sampler_adaptively_2014}, implemented in Stan~\citep{carpenter_stan_2017}.}

\revision{Let $\bstheta^{(r)} = (\bseta^{(r)}, \phi^{(r)})$ denote the $r$-th draw for $r \in \qty[R]$.
These samples are used to impute the post-intervention counterfactual outcomes and estimate treatment effects. 
For each $t > T_0$, we sample and compute:
\begin{equation}
    \label{eq:outcome-sampling}
    \begin{gathered}
        Y_{1t}^{(r)}(0, \mathbf{0}_J) \mid \bsS_t^{(r)}, \bsX_1, \bstheta^{(r)} \sim \Normal(m(\bsS_t^{(r)}, \bsX_1; \bseta^{(r)}), \phi^{(r)}), \\
        \tau_t^{(r)} = Y_{1t}(1, \mathbf{0}_J) - Y_{1t}^{(r)}(0, \mathbf{0}_J),
    \end{gathered}
\end{equation}
where $m(\cdot)$ is the mean function in~\eqref{eq:mean-specification}, the vector $\bsS_t^{(r)}$ contains the previously imputed outcome $Y_{1(t-1)}^{(r)}(0, \mathbf{0}_J)$, and $Y_{1t} \equiv Y_{1t}(1, \mathbf{0}_J)$ is the observed post-intervention outcome.}
This sequential sampling procedure relies on the PPD defined in~\eqref{eq:posterior-predictive-distribution}, whose causal validity follows from Assumptions~1-3.
Pointwise estimates for $\tau_t$ can be easily computed using these draws, such as posterior means and credible intervals~(CIs).

\section{Simulation Studies}
\label{sec:simulations}

\revision{We evaluate the finite sample performance of our proposed priors, defined in~\eqref{eq:distance-horseshoe} and~\eqref{eq:distance-spike-and-slab}, through a series of numerical experiments. 
These scenarios vary the proportion of control units affected by spillover from 0\% to 30\% and estimate the treatment effect $\tau_t$ in~\eqref{eq:target-estimand} at a single post-intervention time point ($t = T_0 + 1$). 
Across all scenarios, we fix the number of pre-intervention periods to $T_0 = 30$ and the number of control units to $J = 50$.}

\revision{We simulate spatial coordinates $\bsP_i \iid \Uniform(0, 1)$ for each unit $i \in \qty[n]$ and define the distance from the treated unit as $d_i = d_P(\bsP_i, \bsP_1)$.
We generate an unobserved covariate $U_i$ and two observed covariates $\bsX_i = (X_{i1}, X_{i2})$, each from the spatially dependent mechanism $(U_i, X_{i1}, X_{i2}) = \psi_i \bsM_i + \bsvarepsilon_i$, where $\bsM_i \iid \MVN(\mathbf{0}_3, 1.25^2 \bsI_3)$ and $\bsvarepsilon_i \iid \MVN(\mathbf{0}_3, 0.1^2 \bsI_3)$.
The spatial weight is defined as $\psi_i = \exp(-d_i^2 / 2)$.
This design induces greater covariate similarity among control units that are closer to the treated unit.}

\revision{Potential outcomes in the absence of intervention follow a three-factor linear model:
\begin{equation*}
    Y_{it}(0, \mathbf{0}_J) = \bsX_i' \bsvartheta + \bsf_t' \bsmu_i + U_i + \epsilon_{it},
\end{equation*}
where $\bsvartheta \in \R^2$ are fixed covariate effects, $\bsf_t \in \R^3$ are time-specific latent factors, $\bsmu_i \in \R^3$ are unit-specific factor loadings, and $\epsilon_{it} \iid \Normal(0, 1.25^2)$ are transitory shocks.
Under the intervention, the observed outcomes for all units incorporate both a spatially decaying spillover effect $\xi_t \in \R$ and the treatment effect $\tau_t = 5$:
\begin{gather*}
    Y_{it} = Y_{it}(0, \mathbf{0}_J) + \xi_{it}(1 - Z_{it}) \one(t > T_0) + \tau_t Z_{it}, \\
    \xi_{it} = \xi_0 \exp(-d_i) \one(d_i < \rho_0),
\end{gather*}
where $\xi_{0t} = -3$ is the baseline spillover effect and $\rho_0 \in \R^+$ controls the proportion of control units exposed to it.}

\revision{We compare our proposed priors (DHS and DS2) with seven alternative methods grouped into two categories: (a)~model-based and (b)~weight-based. 
Model-based approaches include Bayesian structural time series~(BSTS)~\citep{brodersen_inferring_2015} and generalized synthetic control~(GSC)~\citep{xu_generalized_2017} methods, which account for latent structures and time series dynamic.
Weight-based methods include SC using all units (SC-All), a version excluding the four nearest neighbors (SC-NB), and an oracle variant that excludes only the true spillover-affected units (SC-Oracle).
These three weight-based approaches mirror the control selection mechanism in DS2 with $\kappa_d = 0$ and different values of $\rho$.
SC-NB removes nearest neighbors of the treated unit from the donor pool, approximately 8\% of the controls in the simulation studies, which aligns with the number of counties bordering Philadelphia in our data application. 
SC-Oracle excludes units using the true spillover radius $\rho_0$ at each level.
We also consider two penalized regressions as weight-based methods: LASSO~(L1-Reg)~\citep{tibshirani_regression_1996} and ridge~(L2-Reg)~\citep{hoerl_ridge_regression_biased_1970}.
Each regression uses control outcomes as predictors, without a simplex constraint. 
We exclude the standard regression without shrinkage as~$T_0 < J$.}

\revision{We conduct 1,000 independent replications and report relative bias, root mean squared error (RMSE), coverage rate, and average width of 95\% credible or confidence intervals. 
Rejection rates are also reported based on the 95\% CIs for our priors and permutation tests for weight-based methods, following \citet{abadie_synthetic_2010}.
For each control unit, the permutation test reassigns treatment and computes placebo estimates; $p$-values reflect the proportion of control units that, if assumed to be treated, result in estimated effects greater than the observed effect estimate. 
Rejection occurs if $p < 0.05$.}
\revision{Utility scores $u^C_{i,1}$ from~\eqref{eq:utility-function} are calculated for each control unit $i$, using covariate similarity $d_X(\bsX_i, \bsX_1)$ and spatial distance $d_P(\bsP_i, \bsP_1)$. 
We vary the importance weight $\kappa_d \in \qty{0.0, 0.1, 0.5, 1.0}$ and set the DS2 cutoff $\rho$ in~\eqref{eq:distance-spike-and-slab} to exclude approximately 15\% from the donor pool.
Full implementation details for the nine methods are provided in Web~Appendix~C.}

\revision{Overall, our simulations reflect a realistic setting where spillover effects decay with distance and provide strong numerical support for the finite sample performance of our method. 
We emphasize two key points: (i) interference should depend on distance, that is, units farther from the treated unit are less likely to be affected, and (ii) the proportion of spillover-affected controls must be correctly specified. 
The first point underlies both priors and is central to our application; the second pertains specifically to DS2, which requires accurately excluding contaminated units via a binary threshold. 
In contrast, DHS uses continuous shrinkage rather than a hard cutoff, which can result in greater bias than DS2 but may be advantageous when the extent of spillover is uncertain.}

\subsection{\revision{Comparisons to model-based alternatives}}

\begin{figure}[p]
    \centering
    \includegraphics[width=0.8\linewidth]{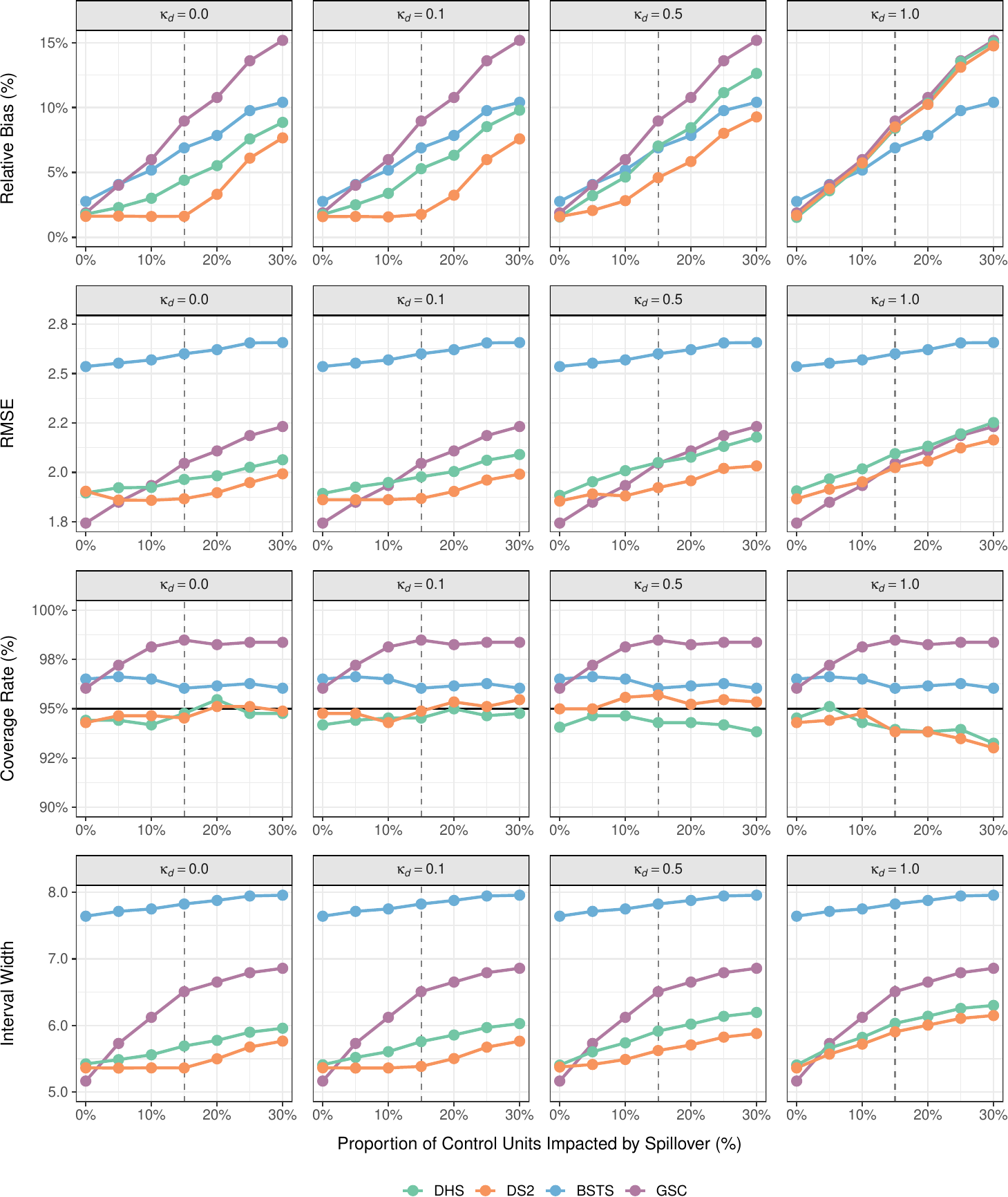}
    \caption{\revision{Performance of our proposed priors (DHS and DS2) compared to (a)~model-based methods (BSTS and GSC) across 1,000 independent replications with $T_0 = 30$ pre-intervention periods and $J = 50$ control units. The panels show, from top to bottom, the relative bias, root mean squared error (RMSE), coverage rate, and average 95\% interval width. Results are reported in increasing proportions of spillover-affected controls (0\% to 30\%). For DHS and DS2, performance is shown at four values of $\kappa_d$, which balances covariate similarity and spatial distance in the utility function $u_C$. Model-based methods are overlaid across all panels, as they do not vary with $\kappa_d$. The dashed vertical line denotes the assumed spillover threshold $\rho$ for DS2; the horizontal line in the coverage panel marks the 95\% nominal level.}}
    \label{fig:simulation-results-a}
\end{figure}

\revision{Figure~\ref{fig:simulation-results-a} demonstrates performance metrics for our proposed priors (DHS and DS2) across varying values of $\kappa_d$, compared to (a) model-based methods (BSTS and GSC), as the proportion of spillover-affected controls increases from 0\% to 30\%. 
When there is no spillover (0\%), all methods perform comparably, with low relative bias and nominal 95\% coverage.
Furthermore, DHS and DS2 remain stable across values of $\kappa_d$, indicating that even without covariate weighting (i.e., when $\kappa_d = 0$), the outcome model in~\eqref{eq:outcome-model} effectively mitigates bias from covariates.}

\revision{As spillover levels rise, bias, interval width, and RMSE increase. 
For our priors, this effect is more pronounced at higher values of $\kappa_d$. 
The increase in interval width for DHS and DS2 is modest and generally stable at varying $\kappa_d$, especially compared to GSC, which exhibits higher inflation and begins to overcover (approximately $98\%$) due to its inability to account for spillover within the bootstrap.
In contrast, DHS and DS2 maintain near nominal coverage at all spillover levels, except when $\kappa_d = 1.0$, where coverage decreases steadily below 95\% as biased estimates from controls affected by spillover shift CIs away from the true $\tau_t$.}

\revision{The stability of DHS and DS2 at $\kappa_d \in \qty{0.0, 0.1}$ suggests that focusing on spatial distance in the utility function helps to exclude contaminated controls, even when ignoring covariate similarity. 
Performance at these values is nearly identical, indicating similar donor pools. 
DS2 consistently achieves the lowest bias due to its hard threshold via $\omega_i$ in~\eqref{eq:distance-spike-and-slab}, which more effectively removes spillover-affected controls than continuous shrinkage through $\lambda_i$ in DHS.
However, when $\rho$ is misspecified below the true threshold (that is, $\rho_0 > \rho$), specifically beyond a spillover level of 15\%, the bias increases.}

\revision{When $\kappa_d = 0.5$, where the spatial distance and covariate similarity are weighted equally, the bias increases for both priors. 
DHS performs comparably with BSTS, while DS2 maintains the lowest bias. 
This increase comes from including controls that are similar in covariates, but spatially nearby and likely affected by spillover. 
Although the baseline spillover effect $\xi_{0t} = -3$ is moderate relative to the standard deviation of the outcomes (1.75), it can introduce substantial bias to affect the results.}
\revision{At $\kappa_d = 1.0$, when selection is based solely on covariate similarity, DHS and DS2 exhibit undesirable performance as the proportion of control units impacted by spillover increases, showing bias comparable or even greater than the comparison methods. 
These results illustrate that considering spatial information in SC selection is critical in the presence of spillover, while remaining bias can be addressed through other components in the outcome model.}

\revision{While BSTS consistently maintains approximately nominal coverage, it produces substantially wider intervals across all spillover levels.
This reflects sensitivity to the hyperparameter that controls uncertainty (variance) around their local trend component (time effect): lower values produce narrower intervals but lead to undercoverage (approximately 85\%), while higher values increase uncertainty and produce nominal coverage at the cost of wider intervals.}

\subsection{\revision{Comparisons to weight-based alternatives}}

\begin{figure}[t]
    \centering
    \includegraphics[width=0.8\linewidth]{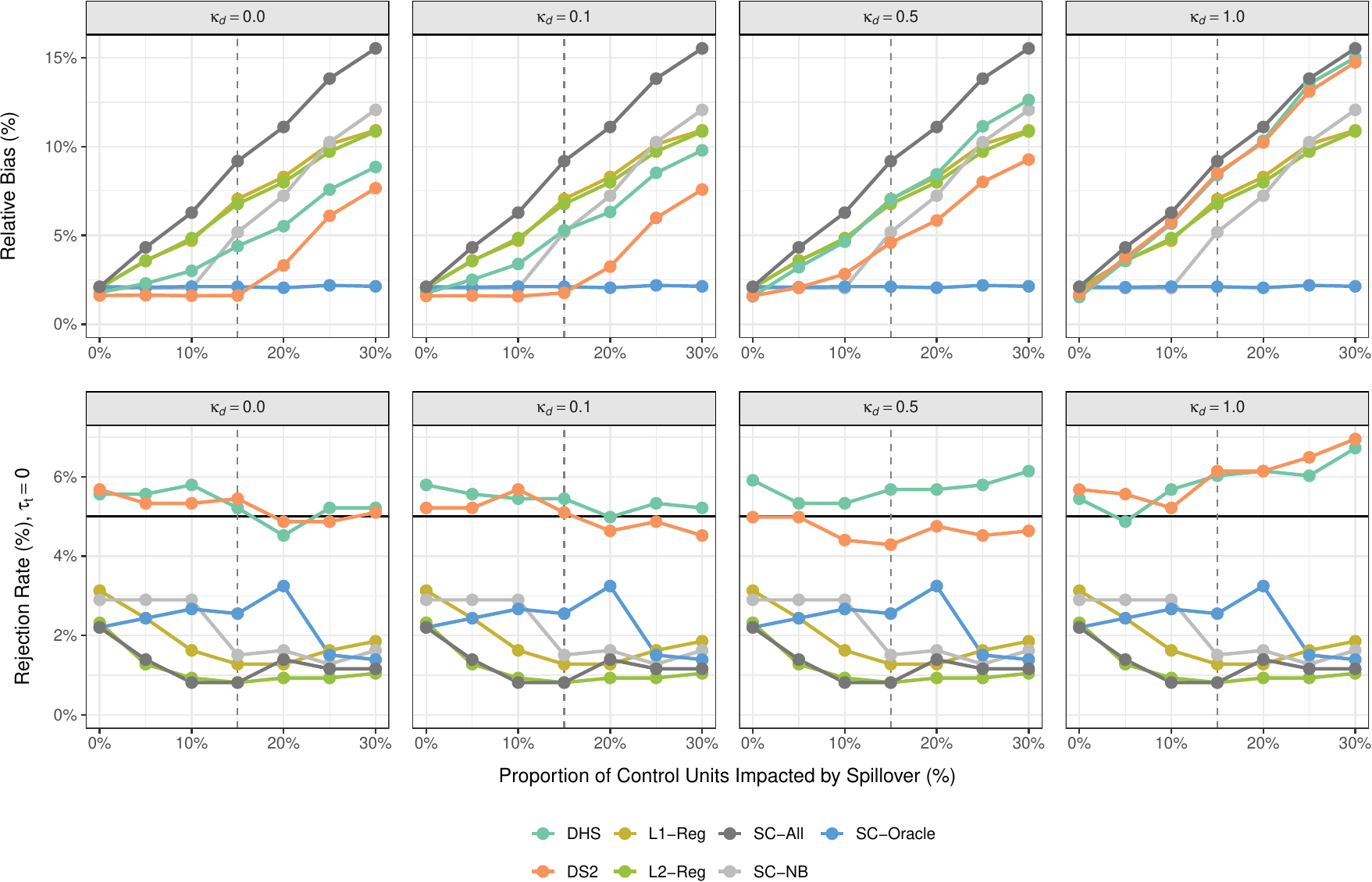}
    \caption{\revision{Performance of our proposed priors (DHS and DS2) compared to (b)~weight-based methods -- L1-Reg, L2-Reg, SC-All, SC-NB, and SC-Oracle -- across 1,000 independent replications with $T_0 = 30$ pre-intervention periods and $J = 50$ control units. Top panel shows relative bias, while the bottom panel shows rejection rates testing a null effect ($\tau_t = 0$). Results are shown in increasing proportions of spillover-affected controls (0\% to 30\%) and varying values of $\kappa_d$ for DHS and DS2. Comparison methods do not depend on $\kappa_d$ and are overlaid across panels. The horizontal line in the rejection panel marks the 5\% nominal level. The dashed vertical line indicates the spillover threshold $\rho$ used in DS2.}}
    \label{fig:simulation-results-b}
\end{figure}

\revision{Figure~\ref{fig:simulation-results-b} shows comparisons between our priors and the five (b) weight-based methods (SC-All, SC-NB, SC-Oracle, L1-Reg, and L2-Reg). 
Performance is evaluated in terms of relative bias and rejection rates under a null effect ($\tau_t = 0$), as confidence intervals are not available for these comparison methods.
For our priors, rejection rates are defined as the proportion of replicates in which the 95\% CIs exclude zero.
Web~Figure~3 further reports rejection rates under a non-zero treatment effect ($\tau_t = 5$).}

\revision{As spillover levels rise, bias increases in all methods that do not exclude affected units. 
For DHS, bias remains lower than that of L1-Reg, L2-Reg, and SC-All at $\kappa_d \in \qty{0, 0.1}$, but increases significantly at $\kappa_d \in \qty{0.5, 1.0}$, performing comparably to those methods due to the introduction of spillover-affected controls.
At $\kappa_d = 0.0$, DS2 performs best in terms of bias when $\rho \geq \rho_0$ excludes all affected units. 
However, when $\rho < \rho_0$, as in the case where the true spillover exceeds 15\%, the bias of DS2 increases, although it still outperforms other methods. 
When $\kappa_d = 0.5$, DS2 remains competitive, but its performance at $\kappa_d = 1.0$ further degrades and is similar to DHS and SC-All.
For methods with selection thresholds (DS2, SC-NB, and SC-Oracle), the bias remains low until their exclusion criteria do not remove all affected units. 
For example, SC-NB begins to degrade beyond 10 to 12.5\% spillover, since it always excludes only four units.
SC-Oracle remains approximately unbiased by design, as it uses the true set of unaffected controls. 
Residual bias in SC-Oracle arises from sample variability and model misspecification.}

\revision{Under $\tau_t = 0$, permutation tests result in lower rejection rates ($< 5\%$).
Even in the absence of spillover, the $p$-values indicate a higher than expected rate of significant results.
In contrast, DHS and DS2 achieve near nominal 5\% rejection rates when $\kappa_d \in \qty{0.0, 0.1, 0.5}$, with slight overrejection at $\kappa_d = 1.0$ as spillover effects can move CIs away from zero.
For SC-NB and SC-Oracle, rejection rates remain stable up to a certain spillover level, after which they decline sharply, likely due to smaller sample sizes from excluding contaminated controls.}

\section{Application to Philadelphia's Beverage Tax}
\label{sec:application}

We apply DHS and DS2, defined in~\eqref{eq:distance-horseshoe} and~\eqref{eq:distance-spike-and-slab}, to evaluate the effect of Philadelphia's beverage tax on SSB purchases in mass merchandise stores.
Sales data are obtained from the NielsenIQ Retail Scanner dataset~\citep{nielseniq_retail_2006}, curated by the Kilts Center at the University of Chicago, which provides detailed weekly pricing and sales volume information across retail chains in the United States.
We focus on mass merchandise stores located in Delaware, Maryland, New Jersey, and Pennsylvania that reported complete sales information on SSB purchased between January~3,~2016, and December~30,~2017.
\revision{The total volume of sales is aggregated at the ZIP3 level from individual store data and then standardized by the number of reporting stores. 
This adjustment ensures comparability across regions, accounting for differences in store density due to population size or varying levels of participation.
As a result, our outcome of interest $Y_{it}$ represents the average sales volume for mass merchandise stores within the ZIP3 region, measured in units of 10,000 fluid ounces.}
Philadelphia is designated as the treated unit and is contained entirely within a single ZIP3 region.
Control units located more than 125 km from Philadelphia are excluded from the analysis to avoid overfitting~\citep{kinn_synthetic_2018}. 

The final data set includes one treated unit and $J = 48$ control units, observed over $T = 26$ time points, with $T_0 = 13$ pre-intervention periods.
The spatial distances $d_P(\bsP_i, \bsP_1)$ are calculated as the Euclidean distance between the geographic centroid of each control unit and that of Philadelphia.
The baseline covariates for each ZIP3 region include demographic composition, income, and population density.
These are sourced from the US Census Bureau and standardized to the unit mean and scale. 
We calculate the \revision{utility scores $u^C_{i,1}$} for each control unit, across $\kappa_d \in \qty{0.0, 0.1, 0.5, 1.0}$.
The DS2 cutoff $\rho$ is set to exclude roughly 25\% of the donor pool.

We estimate the treatment effect $\tau_t$ in~\eqref{eq:target-estimand} for each four-week post-intervention period $t \geq T_0 + 1$.
\revision{For comparison, we also use standard SC with three different donor pools: all control units (SC-All), four bordering units (SC-BR), and non-bordering units (SC-NB). 
These help illustrate how spatial proximity and spillover can influence results.}
\revision{Full implementation details and additional results are provided in Web~Appendix~D.}

\begin{figure}[t]
    \centering
    \includegraphics[width=0.9\linewidth]{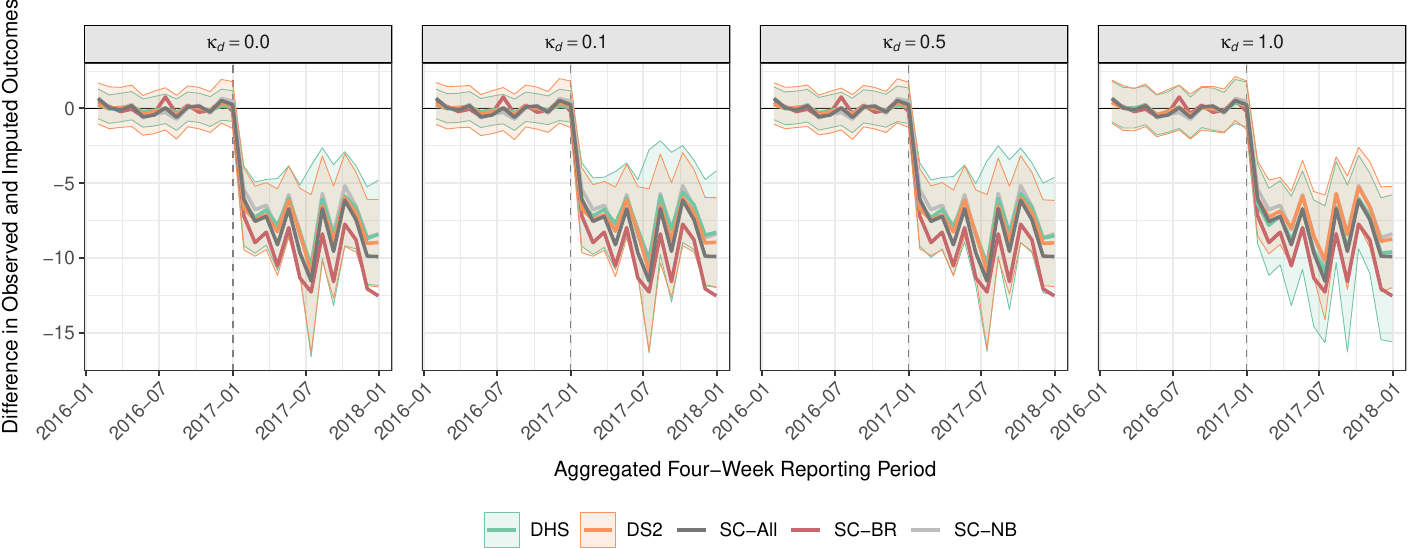}
    \caption{\revision{Posterior mean and 95\% pointwise CIs (displayed as bands) for the differences between observed and imputed beverage sales volume (in units of 10,000 ounces), estimated using DHS and DS2. The results are shown for each four-week period $t \in \qty[T]$ and at varying values of $\kappa_d \in \qty{0.0, 0.1, 0.5, 1.0}$. DS2 uses a cutoff $\rho$ that excludes 25\% from the donor pool. The three SC variants (SC-All, SC-BR, and SC-NB) are overlaid at each panel for comparison. The dashed vertical line marks the tax implementation date; the solid horizontal line indicates a null treatment effect.}}
    \label{fig:application-results}
\end{figure}

Figure~\ref{fig:application-results} shows the posterior mean estimates and 95\% CIs for the difference in observed and imputed outcomes at all time points, separated by $\kappa_d$.
Post-tax estimates of $\tau_t$ show a decrease in sales volume for both priors with CIs that exclude a zero effect, regardless of $\kappa_d$.
Pre-tax differences are centered around zero, indicating a good model fit.
DS2 estimates remain stable across panels, as utility scores include similar control units.
In contrast, DHS estimates become more negative as $\kappa_d$ increases, suggesting that a greater emphasis on covariate similarity may increase the likelihood of including spatially proximate and potentially spillover-affected units, thus overestimating $\tau_t$.
\revision{Web~Table~2 provides the numerical values for Figure~\ref{fig:application-results}.}

\revision{Our data application presents some limitations. 
First, we are restricted to $T_0 = 13$ pre-intervention periods.
Although having more pre-intervention periods can reduce bias, the primary effect of additional data is to decrease uncertainty, leading to narrower CIs. 
Our simulation studies use a larger $T_0$ to illustrate performance under more informative settings.
Second, ZIP3 encompasses larger and more diverse neighborhoods than five-digit ZIP codes (ZIP5).
Aggregating over these broader areas might amplify or diminish spillover effects, affecting the precision of our results.
However, using ZIP5 complicates the assignment of the intervention, since Philadelphia contains several ZIP5 areas. 
This leads to the intervention being assigned to multiple spatially close units, requiring alternative approaches that allow multiple treated units and correlated treatment assignments.
Third, because not all stores across the relevant states are captured in our dataset, we standardize sales volume by the number of reporting stores within each ZIP3 region.
This choice reflects data availability, where population-based standardization would require assuming uniform and complete store coverage, which does not hold in our case.
Instead, we implicitly assume that the ratio of residents to reporting stores is relatively constant between regions.}

\section{Discussion}
\label{sec:discussion}

This work introduces \revision{utility-based} shrinkage priors to estimate treatment effects in the presence of spillover. 
We extend standard Bayesian penalization techniques by developing the DHS and DS2 priors, which data-adaptively select control units for SC in the presence of spillover effects. 
Central to our framework is a \revision{utility function} that governs the degree of shrinkage for each SC coefficient, balancing \revision{covariate similarity} and \revision{spatial distance}.
\revision{Through extensive simulation studies, we demonstrate the effectiveness of our approach compared to seven alternative methods, particularly under spatially dependent confounding and as the level of spillover increases.}

Our framework, which incorporates spatial distance into control selection, is broadly applicable to settings where spillovers may bias treatment effect estimates.
We build on traditional SC methods while introducing key differences.
The classical SC incorporates covariates from all units during the construction of the weights.
However, we include the covariates of the treated unit directly in the outcome model.
Covariates from control units are indirectly used to inform the shrinkage priors without being directly involved in the estimation of SC coefficients.
Furthermore, our utility-based priors offer a novel mechanism to penalize control units based on covariate similarity and spatial distance, features not previously integrated this way.
\revision{Standard penalized SC methods, such as those in \citet{abadie_penalized_2021}, \citet{arkhangelsky_synthetic_2021}, \citet{chernozhukov_exact_robust_conformal_2021},  and \citet{ben-michael_augmented_synthetic_control_2021}, use regularization for variable selection and to avoid overfitting.
More recent Bayesian SC approaches, developed in \citet{brodersen_inferring_2015}, \citet{kim_bayesian_2020}, and \citet{pang_bayesian_2022}, adopt a similar strategy.
Our approach builds on these ideas, but extends them by explicitly addressing spillover-induced bias.}

We acknowledge some limitations.
\revision{First, we recognize the limited exploration of interference structures beyond distance-based spillovers.
Future work could explore alternative interference structures, such as those based on exposure mappings~\citep{aronow_estimating_average_causal_2017}.}
\revision{Second, there are numerous model parameters and hyperparameters that lack a clear guide on how to specify them for applied settings.
We provide practical guidelines for selecting between our priors and setting hyperparameters in Web~Appendix~E.}
Overall, our work seeks to foster discussion and initiate methodological advances in evaluating policy interventions involving controls potentially affected by spillover.

\section*{Acknowledgments}

Researchers' own analyses are derived based in part on data from Nielsen Consumer LLC and marketing databases provided through the NielsenIQ Datasets at the Kilts Center for Marketing Data Center at The University of Chicago Booth School of Business.
The conclusions drawn from the NielsenIQ data are those of the researchers and do not reflect the views of NielsenIQ. NielsenIQ is not responsible, had no role in, and was not involved in analyzing and preparing the results reported here.

\section*{Supplementary Material}

Web Appendices, Tables, and Figures in Sections~\ref{sec:framework}, \ref{sec:methods}, \ref{sec:simulations}, \ref{sec:application}, and \ref{sec:discussion} are available with this article online.
The code used in simulations and data analysis can be found on the GitHub page of the first author at \url{https://github.com/estfernan/Shrinkage-Priors-Spillover-SC}.
An example data set that resembles the structure of the real data used in Section~\ref{sec:application} is provided for illustration purposes, as the original is not publicly accessible.

\section*{Data Availability Statement}

The retail scanner data used in this application are not publicly available but can be accessed upon request through the Kilts Center for Marketing at the University of Chicago Booth School of Business (\url{https://www.chicagobooth.edu/research/kilts/research-data/nielseniq}).

\section*{Funding}

This material is based on work supported by the National Science Foundation Graduate Research Fellowship under Grant No.~2040433.
Any opinions, findings, conclusions, or recommendations expressed in this material are those of the authors and do not necessarily reflect the views of the National Science Foundation.
Funding for Dr. Youjin Lee was provided by the National Institute of Diabetes and Digestive and Kidney Diseases through award R01DK136515.

\bibliography{references.bib}

\clearpage
\newpage

\appendix
\supptitle

\renewcommand{\figurename}{Web Figure}
\renewcommand{\tablename}{Web Table}

\section{\revision{Derivation of Posterior Predictive Distribution}}

\revision{We derive the posterior predictive distribution in a general setting with an arbitrary number of post-intervention periods.
We only need to impute the potential outcomes of the treated unit without intervention; that is, $\bsY^m = \qty{Y_{1t}(0, \mathbf{0}_J) : t > T_0}$.
Therefore, the posterior predictive distribution can be defined as
\begin{equation*}
    p(\bsY^m \mid \bsY^o, \bsZ, \bsV, \bsW)
        \propto p(\bsY^m, \bsY^o, \bsZ, \bsV, \bsW),
\end{equation*}
where we obtain the joint distribution of the completed data.
Let $\bstheta \in \Theta$ and $\bsgamma \in \Gamma$ be a set of parameters, where $\Theta$ and $\Gamma$ are their respective parameter spaces.
Furthermore, let $\bspsi = (\bspsi_Z, \bspsi_V, \bspsi_W) \in \Psi$ denote a set of nuisance parameters, where $\Psi$ is their parameter space.
The parameters $\bstheta$ govern the outcome distribution $Y^m_{1t}(0, \mathbf{0}_J)$, while each element in $\bspsi$ governs the treatment assignments, control outcomes, and covariates, respectively.
We introduce $\bstheta$, $\bsgamma$, and $\bspsi$ into the complete data distribution by marginalizing them out:
\begin{equation*}
    p(\bsY^m, \bsY^o, \bsZ, \bsV, \bsW)
        = \int_\Theta \int_\Gamma \int_\Psi p(\bsY^m, \bsY^o, \bsZ, \bsV, \bsW, \bspsi, \bsgamma, \bstheta) \, \dd \bspsi \, \dd \bsgamma \, \dd \bstheta.
\end{equation*}
We utilize the following factorization:
\begin{align*}
    & p(\bsY^m, \bsY^o, \bsZ, \bsV, \bsW) \\
    & \quad = \int_\Theta \int_\Gamma \int_\Psi p(\bsY^m, \bsY^o, \bsZ, \bsV \mid \bsW, \bspsi, \bsgamma, \bstheta) p(\bstheta \mid \bsW, \bsgamma, \bspsi) p(\bsgamma \mid \bsW, \bspsi) p(\bsW \mid \bspsi)  p(\bspsi) \, \dd \bspsi \, \dd \bsgamma \, \dd \bstheta.
\end{align*}
We can factor the joint distribution of the time-varying components as
\begin{equation*}
    p(\bsY^m, \bsY^o, \bsZ, \bsV \mid \bsW, \bspsi, \bsgamma, \bstheta) 
        = p(\bsY^m, \bsY^o, \bsZ \mid \bsV, \bsW, \bspsi, \bsgamma, \bstheta) p(\bsV \mid \bsW, \bspsi, \bsgamma, \bstheta).
\end{equation*}
Under Assumptions~2-3, the potential outcomes are conditionally independent of the treatment assignments given the control outcomes and covariates; that is, $\bsY^m, \bsY^o \perp \bsZ \mid \bsV, \bsW$~\citep[see][]{gutman_bayesian_procedure_estimating_2018}:
\begin{equation*}
    p(\bsY^m, \bsY^o, \bsZ \mid \bsV, \bsW, \bspsi, \bsgamma, \bstheta)
    = p(\bsZ \mid \bsV, \bsW, \bspsi, \bsgamma, \bstheta) p(\bsY^m, \bsY^o \mid \bsV, \bsW, \bspsi, \bsgamma, \bstheta).
\end{equation*}
Substituting the previous factorization, we obtain
\begin{align*}
    & p(\bsY^m, \bsY^o, \bsZ, \bsV, \bsW) \\
    & \quad = \int_\Theta \int_\Gamma \int_\Psi p(\bsZ \mid \bsV, \bsW, \bspsi, \bsgamma, \bstheta) p(\bsY^m, \bsY^o \mid \bsV, \bsW, \bspsi, \bsgamma, \bstheta) p(\bsV \mid \bsW, \bspsi, \bsgamma, \bstheta) \\
    & \qquad \times p(\bstheta \mid \bsW, \bsgamma, \bspsi) p(\bsgamma \mid \bsW, \bspsi) p(\bsW \mid \bspsi)  p(\bspsi) \, \dd \bspsi \, \dd \bsgamma \, \dd \bstheta.
\end{align*}
Next, we invoke the following assumptions, some derived from the parameter definitions:
\begin{enumerate}
    \item[(a)] The treatment assignments $\bsZ_t$ are solely governed by the parameters $\bspsi_Z$.
    \item[(b)] The outcomes $Y_{1t}(0, \mathbf{0}_J)$ are solely governed by the parameters $\bstheta$.
    \item[(c)] The control outcomes $\bsV_t$ are solely governed by the parameters $\bspsi_V$.
    \item [(d)] The parameters $\bstheta$ are \textit{a priori} independent of $\bsW$ and $\bspsi$, where they are solely governed by $\bsgamma$. 
    \item[(e)] The parameters $\bsgamma$ are \textit{a priori} independent of $\bspsi$ and solely governed by the covariates $\bsW$.
    \item[(f)] The covariates $\bsW$ are solely governed by the parameters $\bspsi_W$.
\end{enumerate}
Applying these independence assumptions, the joint distribution simplifies to
\begin{align*}
    & p(\bsY^m, \bsY^o, \bsZ, \bsV, \bsW) \\
    & \quad = \int_\Theta \int_\Gamma \int_\Psi p(\bsZ \mid \bsV, \bsW, \bspsi_Z) p(\bsY^m, \bsY^o \mid \bsV, \bsW, \bstheta) p(\bsV \mid \bsW, \bspsi_V) \\
    & \qquad \times p(\bstheta \mid \bsgamma) p(\bsgamma \mid \bsW) p(\bsW \mid \bspsi_W)  p(\bspsi) \, \dd \bspsi \, \dd \bsgamma \, \dd \bstheta.
\end{align*}
Since we are only interested in imputing the counterfactual outcomes $\bsY^m$ under the parameters $\bstheta$ and $\bsgamma$, we can absorb the components that do not depend on these quantities into the proportionality constant:
\begin{equation*}
    p(\bsY^m \mid \bsY^o, \bsZ, \bsV, \bsW) 
        \propto \int_\Theta \int_\Gamma p(\bsY^m, \bsY^o \mid \bsV, \bsW, \bstheta) p(\bstheta \mid \bsgamma) p(\bsgamma \mid \bsW) \, \dd \bsgamma \, \dd \bstheta.
\end{equation*}
The joint distribution of the potential outcomes can be factorized over time, yielding a product of conditional distributions for pre- and post-intervention periods:
\begin{align*}
    & p(\bsY^m, \bsY^o \mid \bsV, \bsW, \bstheta) \\
    & \quad = \prod_{t > T_0} p(Y_{1t}(0, \mathbf{0}_J) \mid \obar{Y}_{1(t-1)}(0, \mathbf{0}_J), \bsV, \bsW, \bstheta) \\
    & \qquad \times \prod_{t \leq T_0} p(Y_{1t}(0, \mathbf{0}_J) \mid \obar{Y}_{1(t-1)}(0, \mathbf{0}_J), \bsV, \bsW, \bstheta),
\end{align*}
where, for brevity, we omit the superscripts $m$ and $o$.
Under Assumption~2, each conditional depends solely on the current state vector $\bsS_t = (\bsV_t, Y_{1(t-1)}(0, \mathbf{0}_J))$ and the treated unit's covariates $\bsX_1$.
Hence, for all $t$,
\begin{align*}
    & p(Y_{1t}(0, \mathbf{0}_J) \mid \obar{Y}_{1(t-1)}(0, \mathbf{0}_J), \bsV, \bsW, \bstheta) \\
    & \quad = p(Y_{1t}(0, \mathbf{0}_J) \mid \bsV_t, Y_{1(t-1)}(0, \mathbf{0}_J), \bsW, \bstheta) \\
    & \quad = p(Y_{1t}(0, \mathbf{0}_J) \mid \bsS_t, \bsX_1, \bstheta).
\end{align*}
Combining these results, the posterior predictive distribution of the missing potential outcomes is
\begin{align*}
    & p(\bsY^m \mid \bsY^o, \bsZ, \bsV, \bsW) \\
    & \quad \propto \int_\Theta \int_\Gamma \prod_{t > T_0} p(Y^m_{1t}(0, \mathbf{0}_J) \mid \bsS_t, \bsX_1, \bstheta) \prod_{t \leq T_0} p(Y_{1t}(0, \mathbf{0}_J) \mid \bsS_t, \bsX_1, \bstheta) p(\bstheta \mid \bsgamma) p(\bsgamma \mid \bsW) \, \dd \bsgamma \, \dd \bstheta.\qquad\blacksquare
\end{align*}}

\section{\revision{Technical Details for Our Utility-Based Approach}}
\label{supp:technical-details}

Web~Table~\ref{tab:notations} summarizes the key notations of the distance-based horseshoe (DHS) prior and distance-based spike-and-slab (DS2) prior.
Web~Figures~\ref{fig:dhs-bayesian-network}~and~\ref{fig:ds2-bayesian-network} present the two graphical representations of the outcome model with both priors, respectively.
Refer to the main text for a detailed description of the model and its parameters.
\revision{We provide further details on the model-specific posterior distribution, as well as prior choices for $\bsvartheta$, $\varphi$, and~$\phi$.
Recall that the model for the potential outcomes of the treated unit under no intervention is defined as
\begin{equation*}
    \begin{aligned}
        & \E{Y_{1t}(0, \mathbf{0}_J)}{\bsS_t, \bseta} \\
        & \quad = \bsX_1' \bsvartheta + \bsV_t' \bsbeta + \varphi Y_{1(t-1)}(0, \mathbf{0}_J) \\
        & \quad \eqqcolon m(\bsS_t; \bseta),
    \end{aligned}
\end{equation*}
where $\bseta = (\bsvartheta, \bsbeta, \varphi)$, $\bstheta = (\bseta, \phi)$.
We incorporate covariate effects $\bsvartheta = (\vartheta_1, \ldots, \vartheta_q) \in \R^q$, SC coefficients $\bsbeta = (\beta_2, \ldots, \beta_n) \in \R^J$, an autoregressive coefficient $\varphi \in \R$, and the outcome variance $\phi \in \R^+$.
We assume a normal outcome model, common in Bayesian SC methods~\citep{brodersen_inferring_2015, kim_bayesian_2020, pang_bayesian_2022}, such that $Y_{it}(0, \mathbf{0}_J) \mid \bsS_t, \bseta, \phi \sim \Normal(m(\bsS_t; \bseta), \phi)$. 
For the model parameters, we assign weakly informative independent priors: $\bsvartheta \sim \MVN(\mathbf{0}_q, 3^2 \bsI_q)$, $\varphi \sim \Normal(0, 3^2)$, and $\phi \sim \HS(4, 0, 1)$, where $\HS(4, 0, 1)$ is a positively truncated Student's $t$ distribution with four degrees of freedom and a scale of one.}

\revision{Given the model and prior specifications, the posterior distribution of $\bstheta$ is based on the identification result for the posterior predictive distribution, presented in the main text:
\begin{equation*}
    p(\bstheta \mid \bsY^o, \bsV, \bsW)
        \propto \int_\Gamma \prod_{t \leq T_0} p(Y_{1t}(0, \mathbf{0}_J) \mid \bsS_t, \bstheta) p(\bstheta \mid \bsgamma) p(\bsgamma \mid \bsW) \, \dd \bsgamma,
\end{equation*}
where we marginalize over the data-dependent hyperparameters $\bsgamma$.
We can incorporate each of the individual parameters in $\bstheta$ and invoke the \textit{a priori} independence assumptions imposed by the priors, such that
\begin{align*}
    p(\bstheta \mid \bsY^o, \bsV, \bsW)
        & \propto \int_\Gamma \prod_{t \leq T_0} p(Y_{1t}(0, \mathbf{0}_J) \mid \bsS_t, \bstheta) p(\bstheta \mid \bsgamma) p(\bsgamma \mid \bsW) \, \dd \bsgamma \\
        & \propto \int_\Gamma \prod_{t \leq T_0} p(Y_{1t}(0, \mathbf{0}_J) \mid \bsS_t, \bseta, \phi) p(\bseta, \phi \mid \bsgamma) p(\bsgamma \mid \bsW) \, \dd \bsgamma \\
        & \propto \int_\Gamma \prod_{t \leq T_0} p(Y_{1t}(0, \mathbf{0}_J) \mid \bsS_t, \bseta, \phi) p(\bsvartheta, \bsbeta, \varphi, \phi \mid \bsgamma) p(\bsgamma \mid \bsW) \, \dd \bsgamma \\
        & \propto \int_\Gamma \prod_{t \leq T_0} p(Y_{1t}(0, \mathbf{0}_J) \mid \bsS_t, \bseta, \phi) p(\bsvartheta) p(\varphi) p(\phi) p(\bsbeta \mid \bsgamma, \phi) p(\bsgamma \mid \bsW) \, \dd \bsgamma.
\end{align*}
Density functions can replace the general notation $p(\cdot)$ to facilitate specifying the data likelihood through Stan~\citep{carpenter_stan_2017}.
The choice of shrinkage prior (DHS or DS2) will determine the expressions for $p(\bsbeta \mid \bsgamma, \phi)$ and $p(\bsgamma \mid \bsW)$.}

\revision{DHS places a normal distribution on each SC coefficient $\beta_i$ for $i \in \qty[n] \setminus \qty{1}$:
\begin{equation*}
    \begin{gathered}
        \beta_i \mid \phi, \lambda_i, \zeta \sim \Normal(0, \phi \lambda_i^2 \zeta^2), \\
        \lambda_i \mid u^C_{i,1} \sim \HC(0, u^C_{i,1}), \\
        \zeta \sim \HC(0, 1),
    \end{gathered}
\end{equation*}
where $\zeta \in \R^+$ is the global shrinkage parameter shared across units, $\lambda_i \in \R^+$ is a local parameter that introduces unit-specific variation, and $\phi$ is the outcome variance from the model in the main text.
This formulation adapts the standard horseshoe prior~\citep{carvalho_horseshoe_2010} by embedding the utility score $u^C_{i,1}$ directly into the scale of $\lambda_i$, thus shifting the shrinkage applied to each coefficient based on covariate similarity and spatial distance.
The resulting variance of $\beta_i$ reflects both global regularization through $\zeta$ and local adjustment through $u^C_{i,1}$.
The global shrinkage parameter $\zeta$ follows a weakly informative prior, where $\HC(0, 1)$ is a positively truncated Cauchy distribution with location zero and scale one.}

\revision{DS2, like its traditional counterpart introduced in \citet{mitchell_bayesian_1988} and \citet{george_variable_1993}, imposes a two-component mixture distribution on each SC coefficient $\beta_i$ for $i \in \qty[n] \setminus \qty{1}$:
\begin{equation*}
    \begin{gathered}
        \beta_i \mid \omega_i, \phi, \nu \sim (1 - \omega_i) \delta_0 + \omega_i \Normal(0, \phi \nu^2), \\
        \omega_i \coloneqq \one(u^C_{i,1} > \rho), \\
        \nu \sim \HC(0, 1),
    \end{gathered}
\end{equation*}
where $\omega_i \in \qty{0,1}$ denotes mixture assignment, $\phi \in \R^+$ is the outcome variance from the model in the main text, $\nu \in \R^+$ controls the slab variance, $\delta_0$ is a point mass at zero representing the spike component, and $\rho \in \qty[0, 1]$ is a user-specified cutoff.
This adaptation deterministically sets the mixture assignment based on the utility score $u^C_{i,1}$ and the threshold $\rho$.
The cutoff $\rho$ has different interpretations depending on the value of $\kappa_d$; but the simplest interpretation occurs when $\kappa_d = 0$, where $\rho$ represents a minimum spatial distance.
Control units with a utility score less than $\rho$ are excluded from the SC, whereas those greater than $\rho$ are normally modeled with minimal to no shrinkage.
The slab variance $\nu$ follows the same weakly informative prior as $\zeta$ in the DHS.}

\section{\revision{Additional Simulation Details}}

Recall that the three-factor linear model for the counterfactual outcomes is given by
\begin{equation*}
    \revision{Y_{it}(0, \mathbf{0}_J) = \bsX_i' \bsvartheta + \bsf_t' \bsmu_i + U_i + \epsilon_{it}},
\end{equation*}
where the definitions for the data-generating parameters are given in the main text.
These are pre-specified or generated within each replication from autoregressive processes:
\begin{gather*}
    \bsvartheta = (-0.5, 0.5), \\
    f_{1t} = 0.5 f_{1(t-1)} + \varepsilon_{1t}, \\
    f_{2t} = 1 + 0.5 f_{2(t-1)} + \varepsilon_{2t}, \\
    f_{3t} = 0.5 f_{3(t-1)} + \varepsilon_{3t}, \\
    \mu_{ik} \iid \Uniform(0, 1), \quad \text{for } k = 1, 2, 3, \\
    \bsvarepsilon_t = (\varepsilon_{1t}, \varepsilon_{2t}, \varepsilon_{3t}) \iid \MVN(0_4, I_4).
\end{gather*}
This data-generating process resembles the procedures described in \citet{cao_estimation_2019} and \citet{li_statistical_2020}.

\revision{We implement DHS and DS2 via the Stan probabilistic programming language~\citep{carpenter_stan_2017}, which uses a Hamiltonian Monte Carlo algorithm~\citep{homan_nouturn_sampler_adaptively_2014} to generate posterior samples.
For each replication in the simulation studies, we run a single MCMC chain with 10,000 iterations, discarding the first half as burn-in.
BSTS is implemented using \texttt{CausalImpact} R package~\citep{brodersen_inferring_2015}, with a single MCMC chain of 10,000 iterations, discarding the first half as burn-in and applying the default software settings. 
GSC is implemented using \texttt{gsynth} R package~\citep{xu_generalized_2017}, with cross-validation to select the optimal number of factors, two-way fixed effects, and 1,000 bootstrap replicates to estimate 95\% confidence intervals.
The three SC variants (SC-All, SC-Distant, and SC-Oracle) are implemented by solving the standard SC optimization problem with convex hull constraints. 
We use \texttt{nloptr} R package~\citep{johnson_nlopt_2008}, which interfaces with the NLopt library for nonlinear optimization.
Regularized regression methods (L1-Reg and L2-Reg) are implemented using \texttt{glmnet} R package~\citep{friedman_regularization_paths_generalized_2010}, with default settings.}

\revision{Figure~\ref{fig:simulation-results-c} shows rejection rates under a non-zero treatment effect ($\tau_t = 5$).
When there is no spillover (0\%), all methods exhibit rejection rates close to the 95\% nominal value for $\tau_t = 5$, except L1-Reg, which shows a slightly lower rate.
Rejection rates increase or show more variability for weight-based methods as the spillover increases, due to inflated $p$-values from contaminated placebo distributions.
DHS and DS2 maintain nominal 5\% rejection rates in all scenarios.
For SC-Oracle, the rejection rates remain consistent until the 20\% spillover level.
A similar pattern holds for SC-NB when spillover levels are low ($\leq 10\%$), but as the proportion increases, the methods begin to exhibit non-nominal rejection rates, due to lower sample sizes from excluding contaminated control units.}

\section{Additional Application Details and Results}
\label{supp:application}

Web~Figure~\ref{fig:beverage-sales} presents the time series of SSB volume sales at the unit and aggregate levels. 
For the aggregated time series, units are grouped according to whether they border Philadelphia or are the treated unit itself. 
In general, Philadelphia shows a decrease in relative volume sales following the implementation of the beverage tax on January 1, 2017. 
In contrast, SSB sales for ZIP3 areas bordering Philadelphia increase post-tax, while ZIP3 areas not bordering Philadelphia maintain regular sales, reflecting stationarity.
\revision{Web~Figure~\ref{fig:covariate-spatial-scores} displays covariate similarity and spatial distance for each ZIP3 region, calculated relative to Philadelphia using the utility function $u_C$ defined in the main text.
The left panel ($\kappa_d = 1$) highlights regions with covariate profiles most similar to Philadelphia, including both neighboring areas and Baltimore, a distant control unit unlikely to be affected by spillovers from cross-border shopping. 
The right panel ($\kappa_d = 0$) shows spatial distances, included to confirm the correct implementation of $u_C$.}

\revision{For each prior, we run four independent MCMC chains, each with 10,000 iterations, discarding the first 5,000 as burn-in. 
The baseline covariates are standardized based on their respective mean and variance estimates, while the observed outcomes are standardized using the mean and variance calculated solely from the pre-intervention period.}
\revision{Web~Table~\ref{tab:application-results} presents the posterior mean estimates and 95\% credible intervals for the difference in observed and imputed outcomes.
This table provides the values shown in Figure~4 of the main text.}
\revision{During the pre-tax period, all methods estimate near-zero average differences ($< 0.01$).
After the intervention, all methods estimate negative effects, indicating a reduction in sales volume. 
For DHS, the posterior mean shifts from $-7.48$ to $-8.14$ as $\kappa_d$ increases from 0 to 1, consistent with a potential spillover bias when including neighboring controls (see SC-All and SC-BR).
In contrast, DS2 estimates range from $-7.78$ to $-7.40$, showing less variability due to its binary selection mechanism.}
\revision{SC-NB produces average estimates ($-7.25$) comparable to DHS and DS2.
In contrast, SC-All ($-8.26$) and SC-BR ($-9.81$) produce more negative effects, likely due to the spillover bias of including contaminated controls in the donor pool.
Although bordering units may be similar in covariates, their inclusion can overestimate the treatment effect if cross-border shopping increases adjacent sales, thus exaggerating the decline in Philadelphia.
SC-All is less affected, possibly because unaffected units receive higher weights. 
While we assume that spillovers are most likely in bordering regions, this is supported by the observed increase in average post-tax sales across those areas, as shown in Figure~1 of the main text.}

Web~Figure~\ref{fig:trace-plots} presents trace plots for $\tau_t$ at four randomly selected post-intervention times $t > T_0$, with a single chain randomly selected for DHS and DS2 using $\kappa_d = 0$. 
The trace plots display the post-warm-up period over 5,000 iterations, indicating good mixing and convergence for the Markov chains, which suggests an efficient exploration of the posterior distribution.
Web~Figure~\ref{fig:sensitivity-analysis} presents the estimated causal effect of the beverage tax using DS2 across various cutoff values $\rho$, each representing geographic boundaries as $\kappa = 0$. 
Each cutoff excludes increasingly larger percentages of control units from the donor pool, ranging from 0\% to 50\% in increments of 5\%. 
The pre-intervention fit is omitted to highlight post-intervention estimation differences. 
The results show a reduction in SSB sales, with the $95\%$ CIs indicating a non-zero and negative treatment effect ($\tau_t < 0$). 
Despite an increase in SSB sales for ZIP3 areas bordering Philadelphia (Web~Figure~\ref{fig:beverage-sales}), no clear positive spillover effect is observed, as the estimated reduction of sales increases as more neighboring controls are excluded. 
This pattern may result from the DS2 prior not selecting bordering areas affected by spillover, as other areas may contribute more to the model's predictive power.

\section{\revision{Practical Modeling Recommendations}}

\revision{We provide recommendations to help reduce the challenge of selecting between priors and facilitate broader use of shrinkage-based SC methods in practice. 
We believe this guidance, combined with flexible model structures and strong empirical performance, supports the practical uptake of our approach in real-world policy evaluations.}
\revision{The number of model parameters and hyperparameters introduced by our framework may seem daunting.
To reduce this burden, we recommend using weakly informative priors (Web~Appendix~\ref{supp:technical-details}), which allow the likelihood to dominate the posterior while avoiding implausible estimates.
For users with domain knowledge, these priors can be tailored accordingly.
For example, if the lagged effect $\varphi$ is expected to lie within a plausible range, its prior can be centered around zero with varying levels of dispersion, such as $\Normal(0, 1^2)$ or $\Normal(0, 3^2)$.
Empirical Bayes or other data-driven hyperparameter selection procedures may be useful for future research, but these are beyond the scope of this work.
We emphasize caution when incorporating too much data-driven information into prior distributions, as this risks understating uncertainty.
In particular, since our framework already includes a data-dependent hyperprior.}

\revision{For the importance weight $\kappa_d$, we recommend values in the lower to mid range, such as $\kappa_d \in \qty{0.0, 0.1, 0.5}$.
Our simulations show that these values produce some of the best finite sample properties when spillovers are present.
In contrast, higher values (e.g., $\kappa_d = 1.0$) that ignore spatial distance are more susceptible to spillover bias.
The DS2 threshold $\rho$ is another crucial parameter that affects overall performance.
Simulation results suggest that underestimating $\rho$ -- that is, assuming that fewer controls are affected by spillover than actually are -- leads to higher bias.
Therefore, we recommend setting $\rho$ liberally to exclude a larger portion of the donor pool.
Although this approach may remove some unaffected controls, potentially reducing predictive accuracy and increasing variance, the resulting performance loss should be minimal if the donor pool is sufficiently large.
Analysts can also conduct sensitivity analyses (see~Web~Appendix~\ref{supp:application}) across a range of values for $\rho$, which we encourage in applied settings.}
\revision{Finally, we recommend DS2 when analysts have some insight to guide the specification of $\rho$, or when a hard exclusion rule is desired.
In contrast, we recommend DHS for settings where the spillover structure is more ambiguous. 
DHS does not rely on a binary classification of affected controls, but instead allows continuous shrinkage based on utility scores, introducing uncertainty in the degree of spillover.}

\begin{table}[p]
    \centering
    \caption{Key notations of the distance-based priors described in the main text.}
    \label{tab:notations}
    \begin{adjustbox}{width = \linewidth, center}
        \begin{tabular}{llll}
            \toprule
            Category & Notation & Support & Definition \\ \midrule
            \multirow{4}{*}{Data} & $Y_{1t}(z_{1t}, \mathbf{0}_J)$ & $Y_{1t}(z_{1t}, \mathbf{0}_J) \in \R$ & The potential outcome for the treated unit under $z_{1t} \in \qty{0, 1}$ at time $t$. \\
             & $\bsZ_t = (Z_{1t}, \ldots, Z_{nt})$ & $Z_{it} \in \qty{0, 1}$ & The binary treatment assignments for all units at time $t$. \\
             & $\bsV_t = (Y_{2t}, \ldots, Y_{nt})$ & $Y_{it} \in \R$ & The outcomes for the control units in the donor pool at time $t$. \\
             & $u_{i,1}^C$ & $u_{i,1}^C \in [0, 1]$ & The utility score for control unit $i$. \\ \midrule
            \multirow{4}{*}{Parameters} & $\bsvartheta = (\vartheta_1, \ldots, \vartheta_q)$ & $\vartheta_j \in \R$ & The covariate effects for the treated unit. \\
             & $\bsbeta = (\beta_2, \ldots, \beta_n)$ & $\beta_i \in \R$ & The synthetic control coefficients. \\
             & $\varphi$ & $\varphi \in \R$ & The effect of the previous outcome for the treated unit. \\
             & $\phi$ & $\phi \in \R^+$ & The time-invariant variance for the treated unit outcomes. \\ \midrule
            \multirow{2}{*}{DHS} & $\bslambda = (\lambda_2, \ldots, \lambda_n)$ & $\lambda_i \in \R^+$ & The local shrinkage parameter for $\beta_i$. \\
             & $\zeta$ & $\zeta \in \R^+$ & The global shrinkage parameter for all $\beta_i$'s. \\ \midrule
            \multirow{2}{*}{DS2} & $\bsomega = (\omega_2, \ldots, \omega_n)$ & $\omega_i \in \qty{0, 1}$ & The component assignment for $\beta_i$. \\
             & $\nu$ & $\nu \in \R^+$ & The variance parameter for the normally distributed slab component. \\ \midrule
            \multirow{5}{*}{Others} & $\rho$ & $\rho \in [0, 1]$ & The cutoff for utility-based assignment mechanism. \\
             & $\kappa_d$ & $\kappa_d \in [0, 1]$ & The importance weight for the utility function $u_C(\cdot)$. \\
             & $\delta_0$ & & The point mass at zero. \\
             & $\one(\cdot)$ & & The indicator function. \\
             & $\mathbf{0}_n = (0, \ldots, 0)$ & & The $n$-length zero vector. \\
            \bottomrule
        \end{tabular}
    \end{adjustbox}
    \note[Abbreviations:]{Distance-based horseshoe (DHS) prior; distance-based spike-and-slab (DS2) prior.}
\end{table}

\clearpage
\newpage

\begin{sidewaystable}[p]
    \centering
    \caption{Estimated differences in observed and imputed outcomes before and after the beverage tax.}
    \begin{adjustbox}{width=0.8\linewidth, center}
        
\begin{tabular}[t]{cccccccccccc}
\toprule
 & \multicolumn{4}{c}{DHS} & \multicolumn{4}{c}{DS2} & \multicolumn{3}{c}{SCM} \\
\cmidrule(l{3pt}r{3pt}){2-5} \cmidrule(l{3pt}r{3pt}){6-9} \cmidrule(l{3pt}r{3pt}){10-12}
Reporting Period & $\kappa_d = 0.0$ & $\kappa_d = 0.1$ & $\kappa_d = 0.5$ & $\kappa_d = 1.0$ & $\kappa_d = 0.0$ & $\kappa_d = 0.1$ & $\kappa_d = 0.5$ & $\kappa_d = 1.0$ & All & Bordering & NB \\
\midrule
2016-01-30 & 0.22 & 0.22 & 0.27 & 0.42 & 0.28 & 0.28 & 0.28 & 0.39 & 0.66 & 0.41 & 0.70\\
2016-02-27 & -0.04 & -0.02 & -0.02 & 0.03 & 0.03 & 0.02 & 0.01 & 0.02 & 0.04 & 0.13 & 0.07\\
2016-03-26 & 0.05 & 0.04 & 0.06 & 0.03 & 0.05 & 0.06 & 0.06 & -0.11 & -0.11 & -0.21 & -0.08\\
2016-04-23 & 0.13 & 0.12 & 0.15 & 0.22 & 0.16 & 0.15 & 0.16 & 0.19 & 0.18 & -0.05 & 0.27\\
2016-05-21 & -0.22 & -0.23 & -0.28 & -0.40 & -0.38 & -0.37 & -0.37 & -0.46 & -0.56 & -0.55 & -0.63\\
2016-06-18 & -0.13 & -0.10 & -0.14 & -0.23 & -0.21 & -0.21 & -0.21 & -0.18 & -0.44 & -0.33 & -0.38\\
2016-07-16 & -0.01 & -0.02 & -0.01 & 0.10 & -0.02 & -0.02 & -0.03 & 0.07 & 0.04 & 0.76 & -0.27\\
2016-08-13 & -0.26 & -0.26 & -0.34 & -0.55 & -0.42 & -0.41 & -0.40 & -0.43 & -0.60 & -0.35 & -0.72\\
2016-09-10 & 0.10 & 0.09 & 0.13 & 0.10 & 0.15 & 0.16 & 0.15 & 0.04 & 0.10 & 0.21 & 0.03\\
2016-10-08 & 0.02 & 0.00 & 0.00 & -0.04 & 0.01 & 0.02 & 0.01 & -0.02 & 0.15 & -0.25 & 0.20\\
2016-11-05 & -0.12 & -0.08 & -0.10 & -0.21 & -0.19 & -0.18 & -0.18 & -0.15 & -0.24 & -0.14 & -0.33\\
2016-12-03 & 0.14 & 0.14 & 0.20 & 0.38 & 0.35 & 0.36 & 0.36 & 0.55 & 0.53 & 0.57 & 0.66\\
2016-12-31 & 0.09 & 0.08 & 0.12 & 0.16 & 0.15 & 0.17 & 0.16 & 0.14 & 0.26 & -0.19 & 0.48\\ 
Average & $< 0.01$ & $< 0.01$ & $< 0.01$ & $< 0.01$ & $< 0.01$ & $< 0.01$ & $< 0.01$ & $< 0.01$ & $< 0.01$ & $< 0.01$ & $< 0.01$\\\midrule
2017-01-28 & -6.39 & -6.35 & -6.36 & -6.40 & -6.67 & -6.70 & -6.70 & -6.02 & -6.06 & -7.18 & -5.42\\
2017-02-25 & -7.26 & -7.19 & -7.29 & -7.80 & -7.45 & -7.43 & -7.46 & -7.26 & -7.55 & -8.96 & -6.77\\
2017-03-25 & -6.77 & -6.74 & -6.76 & -7.20 & -7.10 & -7.10 & -7.12 & -6.87 & -7.22 & -8.27 & -6.47\\
2017-04-22 & -7.83 & -7.69 & -7.88 & -8.82 & -8.24 & -8.20 & -8.26 & -8.00 & -9.11 & -10.51 & -8.12\\
2017-05-20 & -5.99 & -5.97 & -6.13 & -6.82 & -6.12 & -6.10 & -6.11 & -5.84 & -6.71 & -7.96 & -5.76\\
2017-06-17 & -8.23 & -8.08 & -8.36 & -9.47 & -8.31 & -8.26 & -8.32 & -8.53 & -9.65 & -11.33 & -8.51\\
2017-07-15 & -10.61 & -10.09 & -10.43 & -10.92 & -11.02 & -10.90 & -10.94 & -10.11 & -11.53 & -12.26 & -10.44\\
2017-08-12 & -6.06 & -5.83 & -6.01 & -6.59 & -6.50 & -6.49 & -6.52 & -5.70 & -6.70 & -8.38 & -5.68\\
2017-09-09 & -8.49 & -8.06 & -8.35 & -9.37 & -8.86 & -8.80 & -8.82 & -8.45 & -9.55 & -11.57 & -8.36\\
2017-10-07 & -5.85 & -5.58 & -5.69 & -6.00 & -5.97 & -5.97 & -5.97 & -5.26 & -6.13 & -7.73 & -5.16\\
2017-11-04 & -6.61 & -6.44 & -6.59 & -7.19 & -6.96 & -6.92 & -6.95 & -6.58 & -7.43 & -8.78 & -6.55\\
2017-12-02 & -8.68 & -8.46 & -8.67 & -9.67 & -9.02 & -8.98 & -9.02 & -8.88 & -9.87 & -12.05 & -8.64\\
2017-12-30 & -8.41 & -8.28 & -8.52 & -9.59 & -8.96 & -8.94 & -8.98 & -8.72 & -9.90 & -12.53 & -8.38\\
Average & $-7.48$ & $-7.29$ & $-7.47$ & $-8.14$ & $-7.78$ & $-7.75$ & $-7.78$ & $-7.40$ & $-8.26$ & $-9.81$ & $-7.25$\\
\bottomrule
\end{tabular}

    \end{adjustbox}
    \note{Posterior mean estimates of DHS and DS2 are reported for each value of $\kappa_d \in \qty{0.0, 0.1, 0.5, 1.0}$. The results are compared to those of traditional SC using three donor pool settings: all control units, the four bordering units, and only non-bordering units (NB). All values are shown in units of 10,000 fluid ounces. The symbol $< 0.01$ indicates average sales volumes below 10,000 fluid ounces. The solid horizontal line separates the pre- and post-tax periods.}
    \label{tab:application-results}
\end{sidewaystable}

\clearpage
\newpage

\begin{figure}[p]
    \centering
    \includegraphics[width=0.75\linewidth]{tikz/dhs.tikz}
    \caption{Graphical representation of DHS. Gray-filled circles indicate observable data, while utility scores used for the data-dependent prior are represented by a square. Parameters comprising DHS are outlined by the black solid line.}
    \label{fig:dhs-bayesian-network}
\end{figure}

\begin{figure}[p]
    \centering
    \includegraphics[width=0.75\linewidth]{tikz/ds2.tikz}
    \caption{Graphical representation of DS2. Gray-filled circles indicate observable data, while utility scores and the cutoff used for the data-dependent prior are represented by a square. Parameters comprising DS2 are outlined by the black solid line.}
    \label{fig:ds2-bayesian-network}
\end{figure}

\clearpage
\newpage

\begin{figure}[p]
    \centering
    \includegraphics[width=\linewidth]{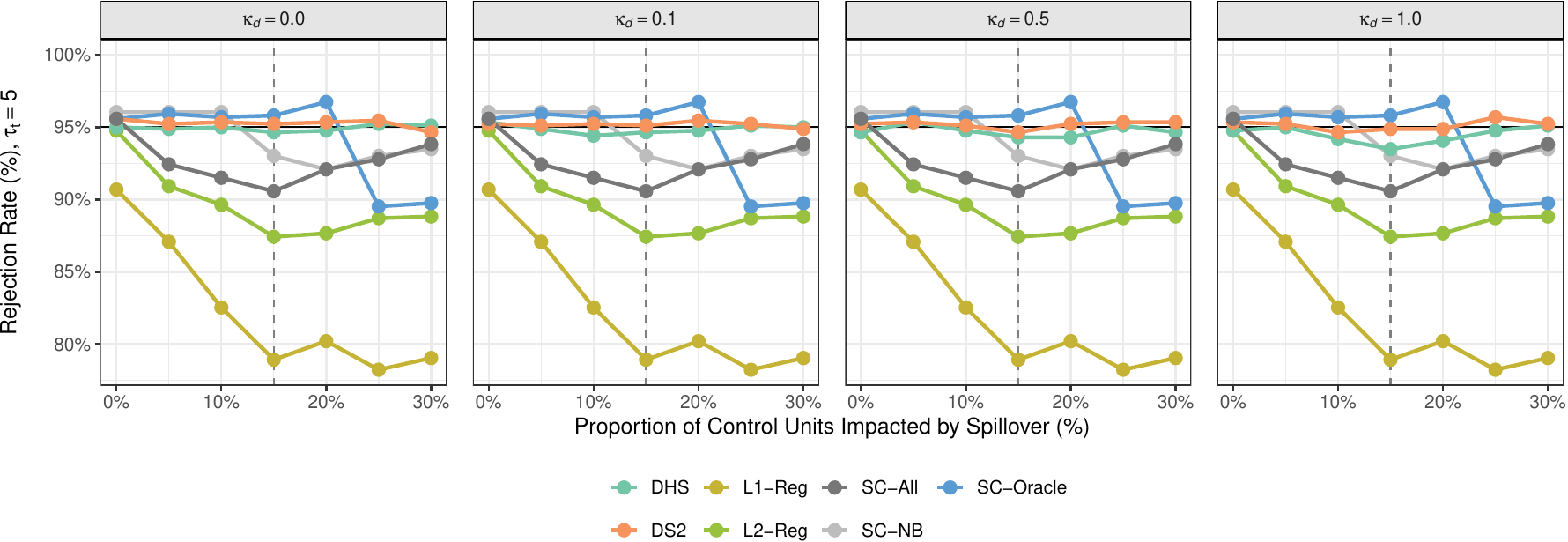}
    \caption{\revision{Rejection rates testing a non-zero effect ($\tau_t = 5$) for our proposed priors (DHS and DS2) compared to (b)~weight-based methods -- L1-Reg, L2-Reg, SC-All, SC-NB, and SC-Oracle -- across 1,000 independent replications with $T_0 = 30$ pre-intervention periods and $J = 50$ control units. Results are shown in increasing proportions of spillover-affected controls (0\% to 30\%) and varying values of $\kappa_d$ for DHS and DS2. Comparison methods do not depend on $\kappa_d$ and are overlaid across panels. The horizontal line in the rejection panel marks the 95\% nominal level. The dashed vertical line indicates the spillover threshold $\rho$ used in DS2.}}
    \label{fig:simulation-results-c}
\end{figure}

\clearpage
\newpage

\begin{figure}[p]
    \centering
    \includegraphics[width=0.75\linewidth]{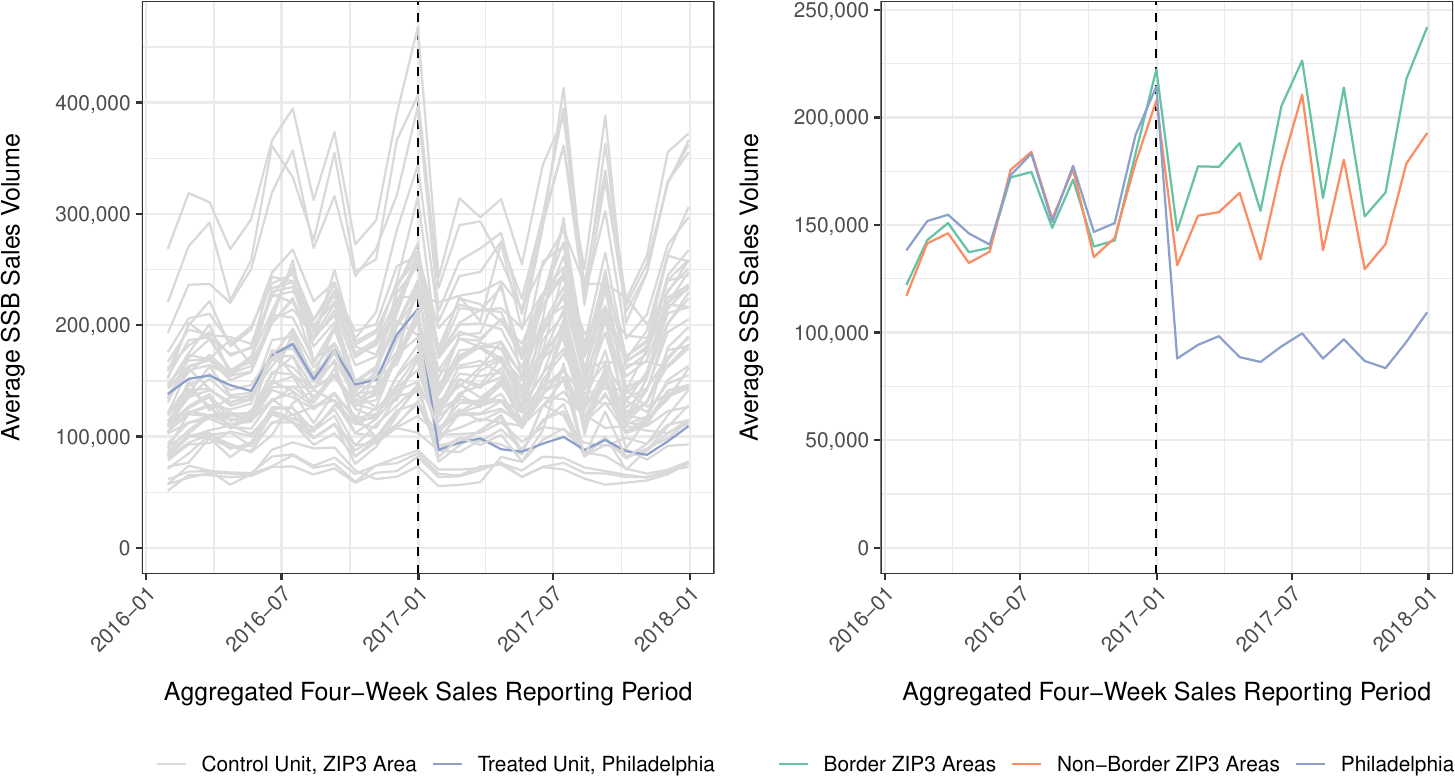}
    \caption{\doublespacing Times-series plots of SSB sales volume, shown at the unit level (left) and aggregated by bordering or treatment status (right). Sales volume is measured from mass merchandise stores within each region and standardized by the number of reporting stores in the ZIP3 area.}
    \label{fig:beverage-sales}
\end{figure}

\begin{figure}[p]
    \centering
    \includegraphics[width=0.75\linewidth]{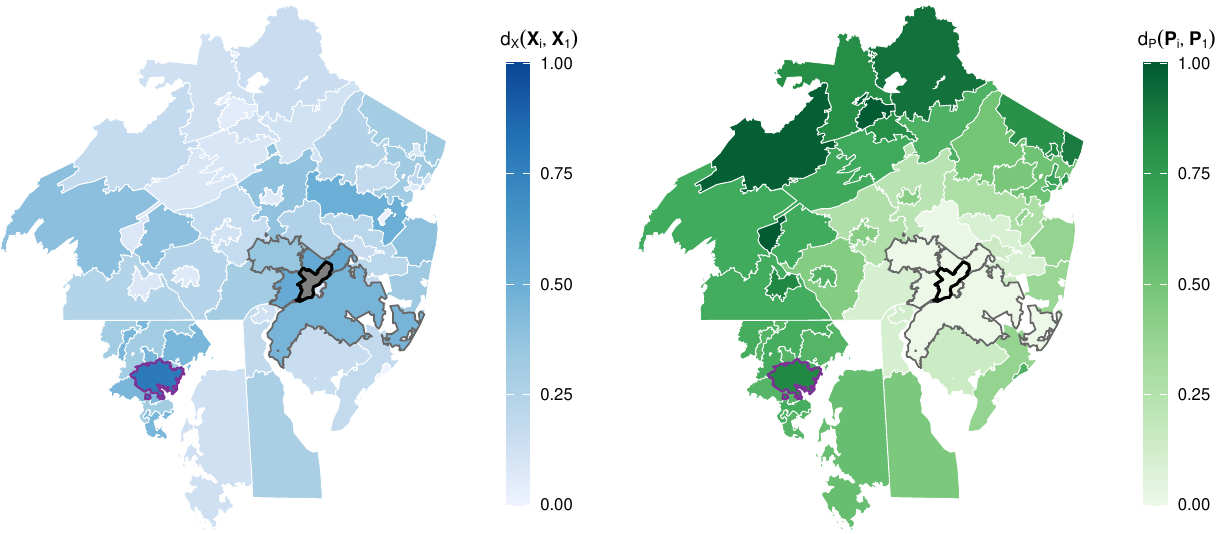}
    \caption{\revision{Covariate similarity and spatial distance for each ZIP3 region, measured relative to Philadelphia using the utility function $u_C$ defined in the main text. Covariate similarity (left panel, $\kappa_d = 1$) reflects pre-tax demographic, income, and population density data from the U.S. Census Bureau, standardized to unit mean and scale. The spatial distance (right panel, $\kappa_d = 0$) is computed using the Euclidean distance between geographic centroids. Key areas are outlined: Philadelphia, the treated unit (black); its bordering regions (gray); and Baltimore (purple), a non-adjacent control with high covariate similarity but minimal spillover risk due to its distance.}}
    \label{fig:covariate-spatial-scores}
\end{figure}

\begin{figure}[p]
    \centering
    \includegraphics[width=\linewidth,height=0.7\textheight,keepaspectratio]{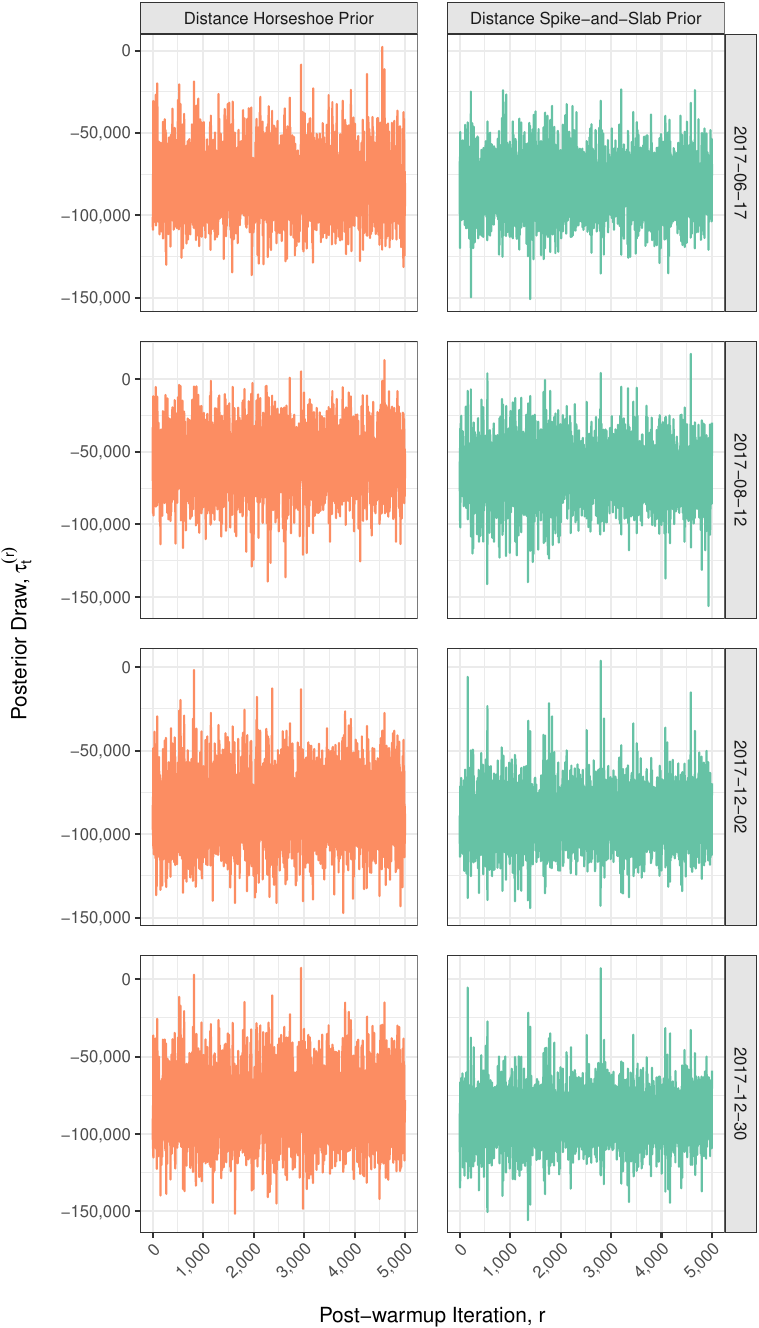}
    \caption{\doublespacing Trace plots for $\tau_t$ at four randomly selected post-intervention times ($t > T_0$) and for a single randomly chosen chain (from 1 to 4) under DHS and DS2 with $\kappa_d = 0$. Each panel shows the post-warm-up period, with the initial 5,000 iterations discarded as burn-in, and the subsequent 5,000 iterations used for inference.}
    \label{fig:trace-plots}
\end{figure}

\begin{figure}[p]
    \centering
    \includegraphics[width=0.7\linewidth,height=0.5\textheight,keepaspectratio]{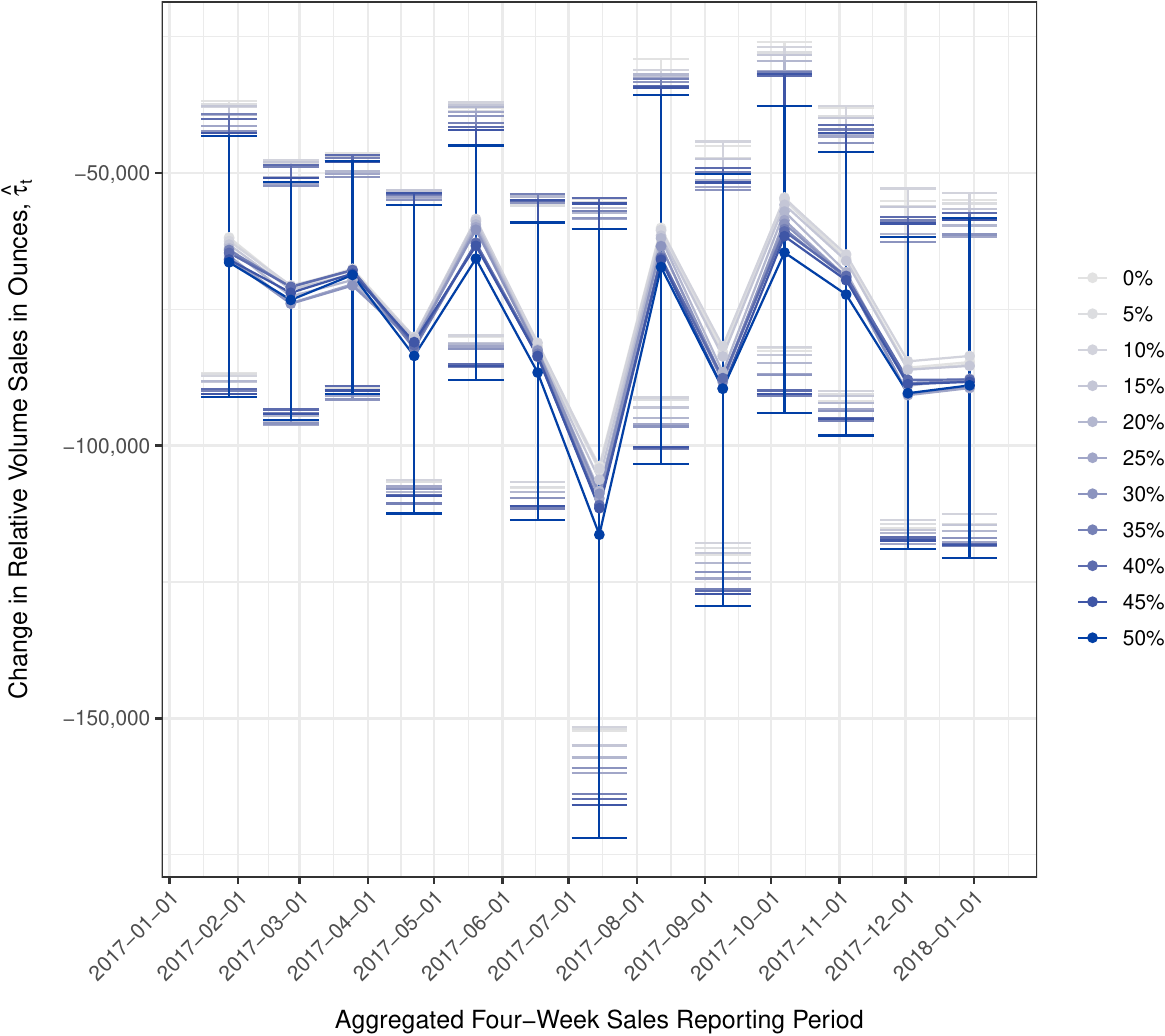}
    \caption{\doublespacing Posterior mean and 95\% CI pointwise estimates for the causal effect $\tau_t$ of the beverage tax on SSB sales in Philadelphia, shown across different assumptions on the percentage of control units affected by spillover for DS2. The cutoff $\rho$ is set according to the sample quantiles of the utility scores, given the percentage, and using $\kappa_d = 0$. Estimates are calculated for each post-intervention four-week aggregated period $t > T_0$.}
    \label{fig:sensitivity-analysis}
\end{figure}

\end{document}